\documentclass[twoside,twocolumn,9pt]{article}

\usepackage{extsizes}
\usepackage[super,sort&compress,comma]{natbib} 
\usepackage[version=3]{mhchem}
\usepackage[left=1.5cm, right=1.5cm, top=1.785cm, bottom=2.0cm]{geometry}
\usepackage{balance}
\usepackage{mathptmx}
\usepackage{sectsty}
\usepackage{graphicx} 
\usepackage{lastpage}
\usepackage[format=plain,justification=justified,singlelinecheck=false,font={stretch=1.125,small,sf},labelfont=bf,labelsep=space]{caption}
\usepackage{float}
\usepackage{fancyhdr}
\usepackage{fnpos}
\usepackage[english]{babel}
\addto{\captionsenglish}{%
  \renewcommand{\refname}{Notes and references}
}
\usepackage{array}
\usepackage{droidsans}
\usepackage{charter}
\usepackage[T1]{fontenc}
\usepackage[usenames,dvipsnames]{xcolor}
\usepackage{setspace}
\usepackage[compact]{titlesec}
\usepackage{xr-hyper} 
\usepackage{hyperref}

\usepackage{epstopdf}

\usepackage{amsmath}
\usepackage{amsfonts}
\DeclareMathOperator*{\argmin}{argmin}
\usepackage{bbold}
\usepackage{siunitx}
\usepackage[ruled]{algorithm2e}
\usepackage{algpseudocode}

\makeatletter
\newcommand*{\addFileDependency}[1]{ 
  \typeout{(#1)}
  \@addtofilelist{#1}
  \IfFileExists{#1}{}{\typeout{No file #1.}}
}
\makeatother

\definecolor{cream}{RGB}{222,217,201}

\begin{document}

\pagestyle{fancy}
\thispagestyle{plain}
\fancypagestyle{plain}{
\renewcommand{\headrulewidth}{0pt}
}

\makeFNbottom
\makeatletter
\renewcommand\LARGE{\@setfontsize\LARGE{15pt}{17}}
\renewcommand\Large{\@setfontsize\Large{12pt}{14}}
\renewcommand\large{\@setfontsize\large{10pt}{12}}
\renewcommand\footnotesize{\@setfontsize\footnotesize{7pt}{10}}
\makeatother

\renewcommand{\thefootnote}{\fnsymbol{footnote}}
\renewcommand\footnoterule{\vspace*{1pt}%
\color{cream}\hrule width 3.5in height 0.4pt \color{black}\vspace*{5pt}} 
\setcounter{secnumdepth}{5}

\makeatletter 
\renewcommand\@biblabel[1]{#1}            
\renewcommand\@makefntext[1]%
{\noindent\makebox[0pt][r]{\@thefnmark\,}#1}
\makeatother 
\renewcommand{\figurename}{\small{Fig.}~}
\sectionfont{\sffamily\Large}
\subsectionfont{\normalsize}
\subsubsectionfont{\bf}
\setstretch{1.125} 
\setlength{\skip\footins}{0.8cm}
\setlength{\footnotesep}{0.25cm}
\setlength{\jot}{10pt}
\titlespacing*{\section}{0pt}{4pt}{4pt}
\titlespacing*{\subsection}{0pt}{15pt}{1pt}

\fancyfoot{}
\fancyfoot[LO,RE]{\vspace{-7.1pt}\includegraphics[height=9pt]{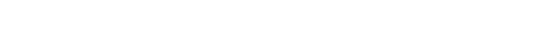}}
\fancyfoot[CO]{\vspace{-7.1pt}\hspace{13.2cm}\includegraphics{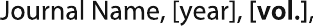}}
\fancyfoot[CE]{\vspace{-7.2pt}\hspace{-14.2cm}\includegraphics{head_foot/RF}}
\fancyfoot[RO]{\footnotesize{\sffamily{1--\pageref{LastPage} ~\textbar  \hspace{2pt}\thepage}}}
\fancyfoot[LE]{\footnotesize{\sffamily{\thepage~\textbar\hspace{3.45cm} 1--\pageref{LastPage}}}}
\fancyhead{}
\renewcommand{\headrulewidth}{0pt} 
\renewcommand{\footrulewidth}{0pt}
\setlength{\arrayrulewidth}{1pt}
\setlength{\columnsep}{6.5mm}
\setlength\bibsep{1pt}

\makeatletter 
\newlength{\figrulesep} 
\setlength{\figrulesep}{0.5\textfloatsep} 

\newcommand{\topfigrule}{\vspace*{-1pt}%
\noindent{\color{cream}\rule[-\figrulesep]{\columnwidth}{1.5pt}} }

\newcommand{\botfigrule}{\vspace*{-2pt}%
\noindent{\color{cream}\rule[\figrulesep]{\columnwidth}{1.5pt}} }

\newcommand{\dblfigrule}{\vspace*{-1pt}%
\noindent{\color{cream}\rule[-\figrulesep]{\textwidth}{1.5pt}} }

\makeatother


\twocolumn[
\begin{@twocolumnfalse}
{
    \includegraphics[height=30pt]{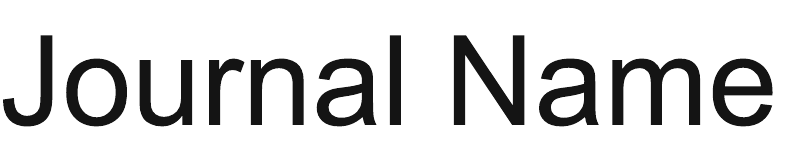}\hfill\raisebox{0pt}[0pt][0pt]{\includegraphics[height=55pt]{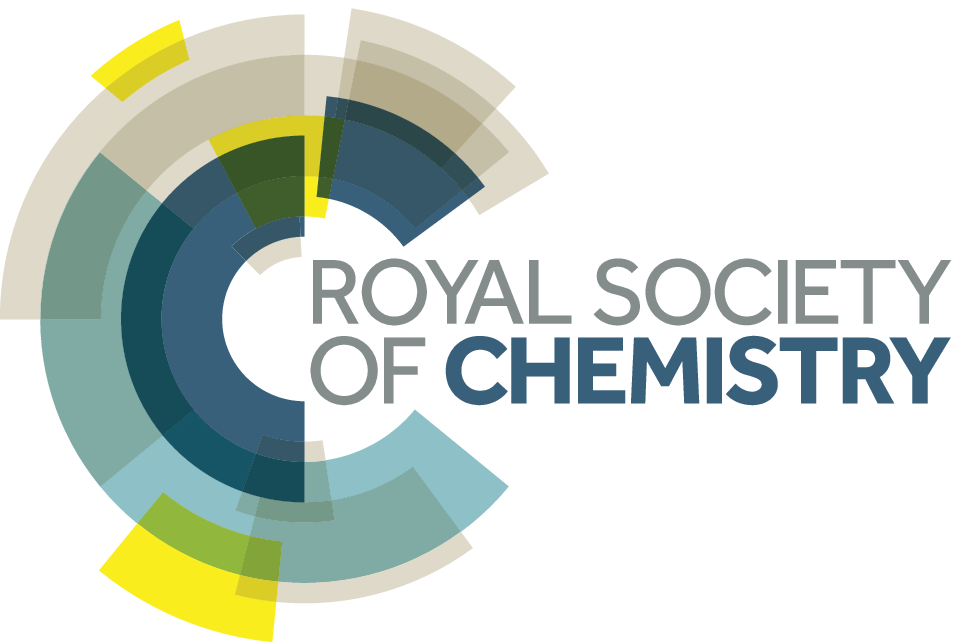}}\\[1ex]
    \includegraphics[width=18.5cm]{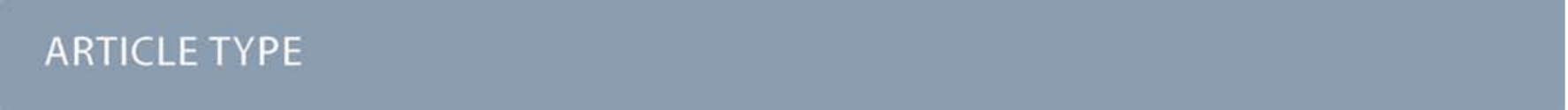}
}\par
\vspace{1em}
\sffamily
\begin{tabular}{m{4.5cm} p{13.5cm} l l}

    \includegraphics{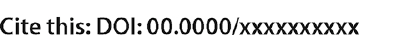} & \noindent\LARGE{\textbf{Generating 3D Molecules Conditional on Receptor Binding Sites with Deep Generative Models}}
    \vspace{0.3cm} & \vspace{0.3cm} \\

    & \noindent\large{Matthew Ragoza,\textit{$^{a}$} Tomohide Masuda,\textit{$^{b}$} and David Ryan Koes\textit{$^{c}$}} \\

    \includegraphics{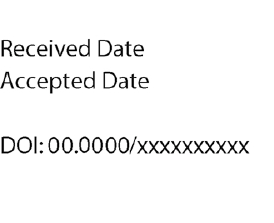} & \noindent\normalsize{The goal of structure-based drug discovery is to find small molecules that bind to a given target protein. Deep learning has been used to generate drug-like molecules with certain cheminformatic properties, but has not yet been applied to generating 3D molecules predicted to bind to proteins by sampling the conditional distribution of protein-ligand binding interactions. In this work, we describe for the first time a deep learning system for generating 3D molecular structures conditioned on a receptor binding site. We approach the problem using a conditional variational autoencoder trained on an atomic density grid representation of cross-docked protein-ligand structures. We apply atom fitting and bond inference procedures to construct valid molecular conformations from generated atomic densities. We evaluate the properties of the generated molecules and demonstrate that they change significantly when conditioned on mutated receptors. We also explore the latent space learned by our generative model using sampling and interpolation techniques. This work opens the door for end-to-end prediction of stable bioactive molecules from protein structures with deep learning.}

\end{tabular}
\end{@twocolumnfalse} \vspace{0.6cm}
]

\renewcommand*\rmdefault{bch}\normalfont\upshape
\rmfamily
\section*{}
\vspace{-1cm}


\footnotetext{\textit{$^{a}$~Intelligent Systems Program, University of Pittsburgh, Pittsburgh, PA, 15213. E-mail: mtr22@pitt.edu}}
\footnotetext{\textit{$^{b}$~Department of Computational and Systems Biology, University of Pittsburgh, Pittsburgh, PA, 15213. E-mail: tmasuda@pitt.edu}}
\footnotetext{\textit{$^{c}$~Department of Computational and Systems Biology, University of Pittsburgh, Pittsburgh, PA, 15213. E-mail: dkoes@pitt.edu}}

\section{Introduction}

Chemical space is enormous, but the subset of molecules that have desirable biological activity is much smaller. Drug discovery typically requires searching through this space for molecules that bind to a specific target, such as a protein implicated in a disease. Thus the search for new drugs involves an alternating procedure of 1) \textit{sampling} compounds from promising regions of chemical space and 2) \textit{screening} them for activity against the biological target. The difficulty of searching chemical space for novel therapeutics has lead to the development of computational methods that sample and screen compounds \textit{in silico} before they are validated experimentally. To reduce the time and cost of drug development, there is growing recognition of the need for new algorithms for sampling compounds with a high chance of success and predicting their biological activity through virtual screening.

Given that the structure of biomolecules determines their function, leveraging the three-dimensional (3D) structure of the target when screening drug candidates has the potential to improve prediction quality. \cite{cheng2012sbvs} The increasing availability of structural data in public repositories like the Protein Data Bank \cite{berman2020pdb} has led to the widespread adoption of machine learning in structure-based drug discovery. Machine learning allows complex nonlinear models of protein-ligand binding to be learned automatically from structural features. Furthermore, the impressive performance of deep learning in computer vision and natural language processing has inspired researchers to apply these methods to biological structures as well. Deep neural networks can learn highly abstract functions from structural data with minimal featurization. This is recently exemplified by AlphaFold, a deep learning model capable of predicting the 3D structures of proteins with high accuracy directly from their amino acid sequence and evolutionary information. \cite{jumper2021alphafold} Despite the utility of deep learning models, it is crucial that they are designed with the appropriate inductive biases and assessed with sufficient cross-validation to avoid overfitting.

Deep learning was first introduced in structure-based drug discovery for scoring the 3D interactions between target proteins (receptors) and small molecules that could potentially bind to them (ligands). Protein-ligand scoring can be formulated as three-dimensional image recognition by training convolutional neural networks (CNNs) on docked protein-ligand poses represented as atomic density grids. This approach has been successfully applied to binding discrimination, \cite{wallach2015atomnet,ragoza2017cnn} pose ranking, \cite{ragoza2017cnn} and affinity prediction.\cite{jimenez2018kdeep, li2019deepatom} Furthermore, grid-based CNN scoring functions have been integrated into ligand pose optimization \cite{ragoza2017opt} for molecular docking, \cite{mcnutt2021gnina} where they outperform traditional scoring functions. Neural networks have also been applied to binding affinity prediction \cite{gomes2017atomconv} and quantum energy estimation\cite{schutt2017schnet} using atomic coordinate-based representations. Deep learning is now widely regarded as the state-of-the-art in virtual screening.

In contrast, it has only recently become viable to use deep learning to sample molecules with drug-like properties prior to virtual screening. Initial efforts to train deep generative models on molecules \cite{gomezbomb2018chemvae, segler2018smilm, ertl2018smilstm} took cues from language modeling by representing molecules with the SMILES string syntax. \cite{weininger1988smiles} Improvements on these approaches used reinforcement learning to guide the generation process towards desired cheminformatic criteria. \cite{olivecrona2017smirl, guimares2018organ} Other work included grammatical constraints that alleviate the tendency for generative models to produce invalid SMILES strings.\cite{kusner2017gramvae, dai2018sdvae} Despite these improvements, SMILES strings are not permutation invariant, so they do not capture the notion of chemical similarity. They also lack conformational information, which  limits their applicability to structure-based drug discovery.

Molecular graphs have been used as a more natural representation of molecules than SMILES strings. Graphs can be provided as input to message-passing neural networks \cite{gilmer2017mpnn} and can be generated as output using fully-connected layers. \cite{simonovsky2018graphvae} However, assuming that generated bonds are independent can result in invalid valences. Solutions include producing molecules as trees of chemically valid substructures \cite{jin2018jtvae} or hard-coding valency constraints into the generative process. \cite{liu2019cgvae, samanta2019nevae} Another concern is that comparing molecules in the loss function requires a graph matching algorithm, which is computationally expensive unless approximations are made. \cite{simonovsky2018graphvae, kwon2019nargvae} Generative adversarial networks (GANs) avoid this by only comparing molecules implicitly, but they are notoriously difficult to train. \cite{decao2018molgan, you2019gcpn} The generation of molecular graphs with deep neural networks can also be biased towards cheminformatic objectives using reinforcement learning. \cite{decao2018molgan, kwon2019nargvae, you2019gcpn}

Most work on deep generative models of molecules have used 2D representations, but contemporary methods can also generate 3D conformations. Early efforts generated different conformers of a single chemical formula using autoregressive models, which output atoms sequentially. \cite{gebauer2018geneq} These have been extended for generating conformers with arbitrary chemical composition \cite{gebauer2020gensym}, simultaneously producing the coordinates and molecular graph, \cite{samanta2019nevae, li2021lnet} and generating linker atoms that connect fragments into valid 3D molecules.\cite{imrie2021linker} Autoregressive models can be made invariant to rotations and translations by modeling distributions over interatomic distances instead of coordinates. However, they require selecting a canonical atomic ordering due to lack of permutation invariance. On the other hand, non-autoregressive approaches generate distance matrices all at once, \cite{hoffmann2019euclid} and have been extended to generating conformers conditioned on a molecular graph. \cite{mansimov2019molgeo, simm2019distgeo} One challenge of non-autoregressive models is that generating distance matrices requires enforcing the triangle inequality. Another drawback is that the Hungarian algorithm, which has cubic time complexity in the number of atoms, must be applied to compare distance matrices in a permutation agnostic manner.

Atomic density grids can also be used as a 3D representation of molecules for generative modeling. Unlike distance matrices, grids are coordinate frame-dependent. However, they are permutation invariant and can be compared without expensive matching algorithms. Density grids also provide holistic shape information that is not easily accessible from atomistic representations, and is arguably of equal importance for protein-ligand binding as pairwise interactions. The main obstacle to generative modeling with atomic density grids is converting them into discrete molecules. Past work has used Wiener deconvolution to approximate the inverse of the density kernel,\cite{kuzminykh2018waveae} but this does not lead to an unambiguous set of atoms and bonds. Another group trained an auxiliary captioning network to output SMILES strings based on density grids, \cite{skalic2019shapevae} but this relinquishes the 3D structure generated by the model. Iterative atom fitting and bond inference is the only approach, to our knowledge, that produces 3D molecular structures from atomic density grids.\cite{ragoza2020gen}

The utilization of protein structure to generate molecules with deep learning is presently an under-explored research area. Preliminary work has generated SMILES strings based on receptor binding site information represented as atomic density grids \cite{skalic2019shapecgan} or Coulomb matrices.\cite{xu2020recsmi} Other researchers used reinforcement learning to guide the sampling of 3D ligands from a generative model towards high predicted affinity for a target protein.\cite{li2021lnet} However, generating 3D molecular structures directly from protein binding pockets remains an unsolved challenge.\cite{masuda2020gen} To address this, we make the following contributions in this work:

\begin{enumerate}
    \item The first demonstration of 3D molecular structure generation with receptor-conditional deep generative models.
    \item Evaluation of the effect on generated molecules of conditioning the generative model on mutated receptors.
    \item Exploration of the latent space learned by the generative model through sampling and interpolation.
\end{enumerate}

\section{Methods}

\begin{figure}
 \centering
 \includegraphics[width=0.8\columnwidth]{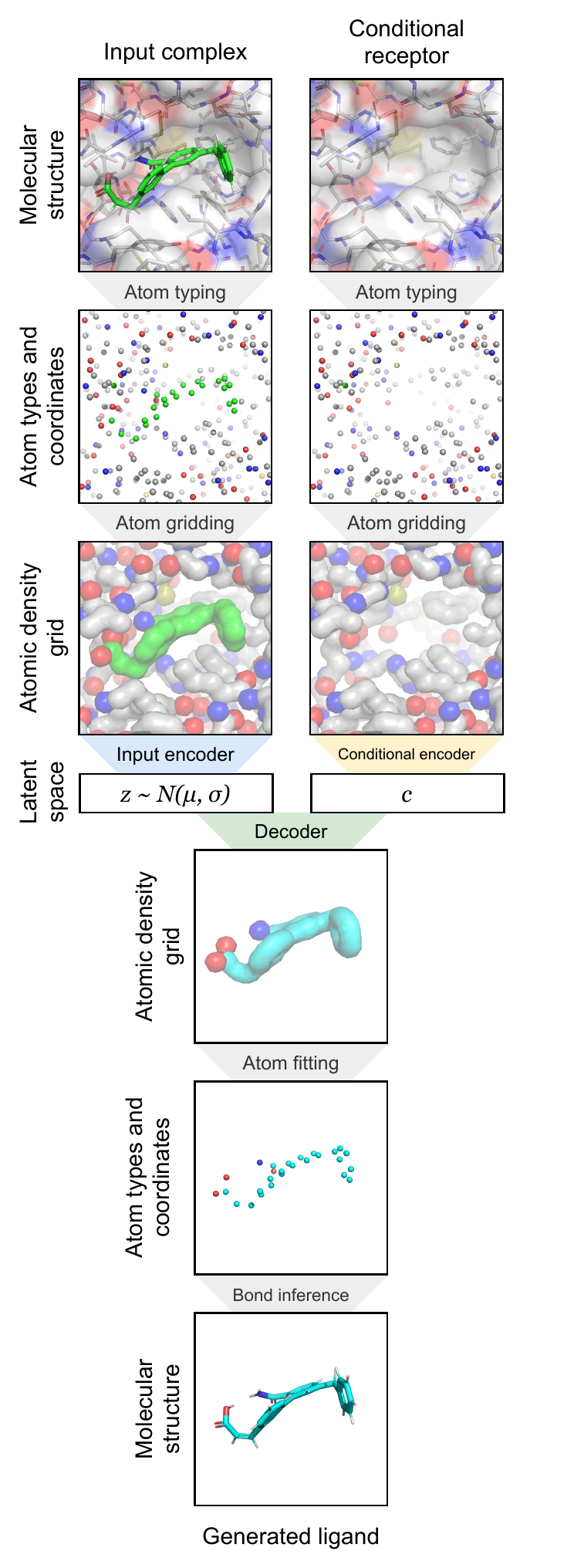}
 \caption{\textbf{Overview of generative modeling pipeline.} First, a docked protein-ligand complex is converted to an atomic density representation through atom typing and gridding operations. Density grids are then provided as input to a conditional variational autoencoder (CVAE). The CVAE input branch encodes the full complex density, while its conditional branch encodes only the receptor density. The complex density is mapped to a probabilistic latent space, which is then sampled as a latent vector $z \sim N(\mu,\sigma)$. This is combined with the conditional vector $c$ output by the conditional encoder, and together they are provided to the decoder. The decoder generates an output ligand density that is converted into the final 3D molecular structure through atom fitting and bond adding.}
 \label{fig:cond_gen_methods}
\end{figure}

\subsection{Property-based atom typing}

First, we assign atom types to molecules using a set of $N_p$ atomic property functions $p$ and value ranges for those properties $\mathbf{v}$, which are listed in Table~\ref{tab:atom_typing}. For a given atom $a$, the atom type vector $\mathbf{t} \in \mathbb{R}^{N_T}$ is created by concatenating $N_p$ atomic property vectors $\mathbf{p}$ through the following:
\begin{align}
\begin{split}
    \mathbf{t}(a) &= \left[ \mathbf{p}(a, (p, \mathbf{v})_i), \ldots, \mathbf{p}(a, (p, \mathbf{v})_{N_p}) \right] \\
    \mathbf{p}(a, (p, \mathbf{v}))_i &= \mathbb{1}(p(a) = \mathbf{v}_{i})
\end{split}
\end{align}

The atomic properties we used were element, aromaticity, H-bond donor and acceptor status, and formal charge. Different element ranges were represented for receptor atoms and ligand atoms, but the value ranges for all other properties were the same. The process we used to construct value ranges for properties and compare different type schemes is described in the supplement.

\begin{table}[h]
\small
    \begin{tabular*}{\columnwidth}{@{\extracolsep{\fill}}lll}
    \hline
        Atomic property & Value range & Num. values \\
        \hline
        Ligand element & B, C, N, O, F, P, S, Cl, Br, I, Fe & 11 \\
        Receptor element & C, N, O, Na, Mg, P, S, Cl, K, Ca, Zn & 11 \\
        Aromatic & False, True & 2 \\
        H-bond acceptor & True & 1 \\
        H-bond donor & True & 1 \\
        Formal charge & -1, 0, 1 & 3 \\
        \hline
    \end{tabular*}
    \caption{\textbf{Atom typing scheme.} The atomic properties and associated ranges of values that were represented in our atom type vectors.}
    \label{tab:atom_typing}
\end{table}


\subsection{Atomic density grids}

After assigning atom type vectors, we convert molecules to an atomic density grid format. Atoms are each represented as continuous densities with a truncated Gaussian shape. The density value of an atom at a grid point is defined by a \textit{kernel} function $f: \mathbb{R} \times \mathbb{R} \rightarrow \mathbb{R}$ that takes as input the distance $d$ between the atom coordinate and the grid point and the atomic radius $r$:

\begin{equation}
    f(d,r) = \\
    \begin{cases} 
      e^{-2(\frac{d}{r})^2} & d \leq 1.5r \\
      0 & d > 1.5r 
   \end{cases}
\end{equation}

The radius was fixed at $r=1.0$ for all atoms in this work. Grid values are computed by summing the density kernel of each atom at each point on a 3D grid, multiplied by the value of the atom's type vector in the corresponding grid channel. A molecule with $N$ atoms and atom type vectors of length $N_T$ can be represented as a matrix of atom types $T \in \mathbb{R}^{N \times N_T}$ and a matrix of atomic coordinates $C \in \mathbb{R}^{N \times 3}$. The function that computes atomic density grids $g: \mathbb{R}^{N \times N_T} \times \mathbb{R}^{N \times 3} \rightarrow \mathbb{R}^{N_T \times N_X \times N_Y \times N_Z}$ is then defined as follows:

\begin{equation}
    \label{eqn:atom_gridding}
    g(T,C)_{txyz} = \sum_{a=1}^{N} T_{at} f(\|C_i-s(x,y,z)\|, 1.0)
\end{equation}

$N$ is the number of atoms in the input molecule, so there is no maximum number of atoms per grid. All atoms that fit within the spatial extent of the grid are represented. We used cubic grids with side lengths of \SI{23.5}{\angstrom} and \SI{0.5}{\angstrom} resolution, resulting in spatial dimensions $N_X = N_Y = N_Z = 48$. Equation~\ref{eqn:atom_gridding} requires a coordinate frame mapping $s: \mathbb{Z}^3 \rightarrow \mathbb{R}^3 $ from grid indices to spatial coordinates. We center the grids on the input molecule before adding random translations and rotations during both training and evaluation. This is facilitated by computing grids on-the-fly using \texttt{libmolgrid}, a GPU-accelerated molecular gridding library. \cite{sunseri2020molgrid}

\subsection{Atom fitting algorithm}

The inverse problem of converting a reference density grid $\mathbf{G}_{ref}$ into a discrete 3D molecular structure does not
have an analytic solution, so we solve it as the following optimization problem:
\begin{equation}
    T^*, C^* = \argmin_{T,C} \| \mathbf{G}_{ref} - g(T, C) \|^2
\end{equation}
 We can detect initial locations of atoms on a grid by selecting from the grid points with the largest density values. \texttt{libmolgrid} allows us to compute the grid representation of an atomic structure and backpropagate a gradient from grid values to atomic coordinates. Therefore, we devised an algorithm that combines iterative atom detection with gradient descent to find a set of atoms that best fits a reference density, shown in Algorithm~S\ref{alg:atom_fitting}.

\subsection{Bond inference algorithm}

We construct valid molecules from the sets of atoms detected by atom fitting using a sequence of inference rules that add bond information and hydrogens while trying to satisfy the constraints defined by the atom types. The algorithm is based on customized bond perception routines implemented in OpenBabel.\cite{openbabel,oboyle2011openbabel} An overview of the procedure is shown in Algorithm~S\ref{alg:bond_adding}.

\subsection{Conditional variational autoencoder}

Our generative model is a conditional variational autoencoder (CVAE) \cite{sohn2015cvae} of atomic density grids. The objective is to learn to sample from the distribution $p(lig|rec)$, where $rec$ is a receptor binding site density and $lig$ is the density of a ligand that binds to it. We assume that there is a latent variable $z$ representing binding interactions that follows a prior distribution we can sample, such as a standard normal distribution. The generative process consists of drawing a sample $z \sim p(z)$ followed by $lig_{gen} \sim p_{\theta}(lig|z,c)$, where $p_\theta$ is a $decoder$ neural network and $c$ is an encoding of a receptor density $rec$ produced by a \textit{conditional encoder} network. Training this model by naive maximum likelihood estimation would require computing the latent posterior probability $p_{\theta}(z|rec,lig)$, which is intractable. The key is to instead train an \textit{input encoder} network to learn an approximate model $q_{\phi}(z|rec,lig)$ of the posterior distribution. The training task minimizes two objectives:
\begin{align}
    L_{recon} & = -\log p_{\theta}(lig|z,c) \propto \frac{1}{2} \|lig - lig_{gen} \|^2\\
    L_{KL} &= D_{KL}(q_{\phi}(z|lig,c)||p(z))
\end{align}

The reconstruction loss term $L_{recon}$ term maximizes the probability that latent samples from the approximate posterior distribution $z \sim q_{\phi}(z|rec,lig)$ are decoded as realistic ligand densities--specifically, the real ligand density $lig$ that was provided to the input encoder. The Kullback-Liebler divergence term $L_{KL}$ encourages the approximate posterior distribution to match the true prior distribution, $p(z) = N(0,1)$. The combined effect is that the latent space follows a normal distribution, enabling generative sampling, while the decoded samples are expected to appear realistic in the receptor context. The model is trained by providing real $(rec, lig)$ examples to the encoder to get latent representations of their interactions, then maximizing the likelihood of decoding the latent vectors back to the corresponding ligand densities when conditioned on the cognate receptor density.

\paragraph*{Steric clash loss.} We included an additional term in the loss function that minimized steric clash in terms of the overlap between the generated ligand density and the receptor density. This was calculated by first summing across the grid channels, then multiplying the receptor and ligand density at each point:
\begin{equation}
    L_{steric} = \langle \sum_i^{N_T} rec_i, \sum_i^{N_T} lig_{gen,i} \rangle \\
\end{equation}

We validated this as a measure of steric clash by checking empirically that real protein-ligand complexes did not have density overlap, owing to our use of a density kernel with a relatively small, fixed atomic radius. We combined the three loss terms with weights into the final loss function like so:
\begin{equation}
    L = \lambda_{recon} L_{recon} + \lambda_{KL} L_{KL} + \lambda_{steric} L_{steric}
\end{equation}

The loss weights were initialized at $\lambda_{recon} = 4.0$, $\lambda_{KL}=0.1$, and $\lambda_{steric} = 1.0$, though the KL divergence loss weight was gradually ramped up to 1.6 over 200,000 iterations, starting at iteration 450,000. The model was trained using RMSprop with learning rate $1e-5$ for 1,000,000 iterations with a batch size of 8.

\begin{figure}[t]
 \centering
 \includegraphics[width=0.75\columnwidth]{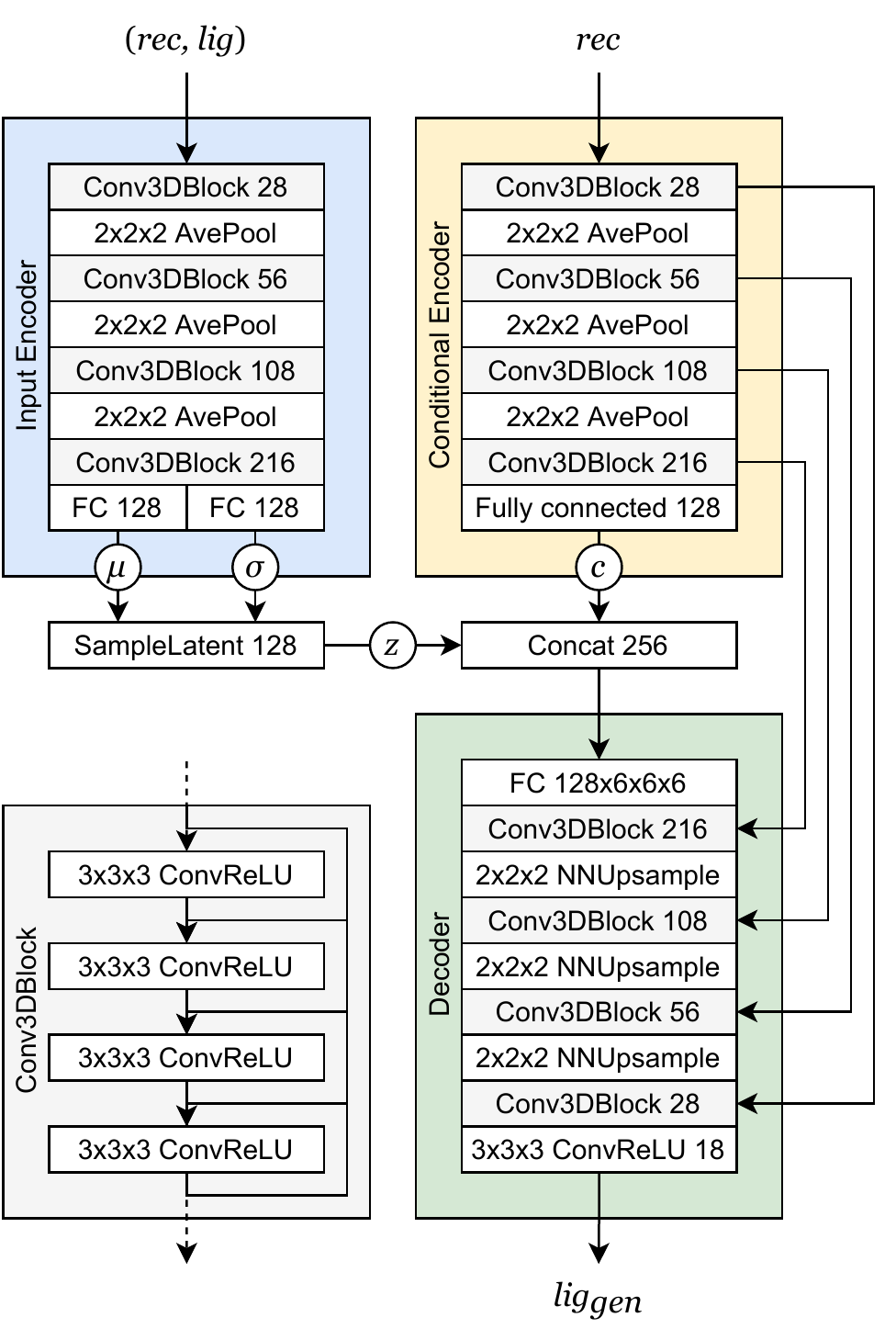}
 \caption{\textbf{Generative model architecture.} The input encoder maps a protein-ligand complex to a set of means and standard deviations defining latent variables, which are sampled to produce a latent vector $z$. The conditional encoder maps a receptor to a conditional encoding vector $c$. The latent vector and conditional vector are concatenated and provided to the decoder, which maps them to a generated ligand density grid. The input encoder and conditional encoder consist of 3D convolutional blocks with leaky ReLU activation functions and residual connections\cite{he2015deep} (see detail of Conv3DBlock), alternated with average pooling. The decoder uses a similar architecture in reverse, with transposed convolutions and nearest-neighbor upsampling instead of pooling. U-Net skip connections\cite{ronneberger2015unet} were included between the convolutional features of the conditional encoder and the decoder to enhance the processing of receptor context. Spectral normalization\cite{miyato2018spectral} was applied to all learnable parameters during training. The value displayed after module names in the diagram indicates the number of outputs (or feature maps, for convolutional modules). If not specified, the number of outputs did not change from the previous layer.}
 \label{fig:cond_gen_arch}
\end{figure}

\subsection{Training data set}

The CrossDocked2020 data set is a massive collection of small molecules docked into cognate and non-cognate receptors.\cite{francoeur2020crossdock} An initial set of 18,450 bound protein-ligand crystal structures were clustered by pocket similarity and then input to a combinatorial docking procedure. Each ligand was re-docked to its known receptor and cross-docked to every other receptor with a similar pocket. Though noisier than crystallized or re-docked poses, cross-docking greatly increases the amount of training data and captures the distribution of structures that are typically used in drug discovery. The CrossDocked2020 data set has cross-validation splits based on pocket similarity. We used the first split to construct our training and test data sets. We omitted any poses that had root-mean-squared deviation (RMSD) greater than \SI{2}{\angstrom} from the crystal pose of the ligand in its cognate receptor. We also omitted molecules that could not be sanitized with RDkit.\cite{rdkit}

\subsection{Test target selection}

We randomly selected ten targets from the CrossDocked2020 test set to evaluate our model. Each target came from a different pocket cluster, and we only considered targets with at least five unique ligands. We used the top-ranked docked pose of each ligand in the set. The test targets and ligands are shown in Table~\ref{tab:test_targets}.

\begin{table}[h]
\small
  \begin{tabular*}{0.48\textwidth}{@{\extracolsep{\fill}}lll}
    \hline
    PBD ID & Ligand IDs & Num. ligands \\
    \hline
    2ah9 & bgn, udp, udh, cto, ud2, upg  & 6 \\
    5lvq & aly, 5wv, 5wz, 2lx, 5ws, 5wu, 2qc, 78y, & 14 \\
         & 5wy, 5x0, 5wt, p2l, 82i, 5wx & \\
    5g3n & x28, oap, 8in, 6in, u8d, bhp, i3n, gel & 8 \\
    1u0f & g6p, 6pg, s6p, der, f6p, a5p & 6 \\
    4bnw & 36k, nkh, 36i, j2t, fxe, q7u, 3x3, 9kq, 36p,  & 13 \\
         & 8m5, 34x, 36e, 36g & \\
    4i91 & cpz, 85d, cae, sne, tmh, 3v4, 82s & 7 \\
    2ati & avf, ave, ihu, 055, 25d, mrd, avd & 7 \\
    2hw1 & tr4, lj9, a4j, tr2, anp, a4g, a3y, a3j, quz, & 11 \\
         & a1y, a2j & \\
    1bvr & xt5, tcu, 3kx, 3ky, 2tk, i4i, uud, geq, 665, & 11 \\
         & nai, nad & \\
    1zyu & adp, skm, anp, acp, s3p, dhk, k2q & 7 \\

    \hline
  \end{tabular*}
  \caption{\textbf{Test set targets.} The proteins that were selected for test evaluations and the associated ligands that were docked to them. Each of the test set proteins has a binding site from a different pocket cluster.}
  \label{tab:test_targets}
\end{table}

\subsection{Sampling methods}

Our model has two distinct sampling modes called \textit{posterior} and \textit{prior} sampling. The difference is whether the generative process is biased towards a particular real protein-ligand interaction, or if it is only based on the conditional receptor. With posterior sampling, a real protein-ligand complex is encoded into the latent variable parameters before drawing samples. In contrast, prior sampling draws latent vectors from a standard normal distribution, so it has no intentional bias towards a specific real ligand. Using either method, the latent vectors are combined with the conditional receptor encoding before decoding an output ligand density. The known ligand for the conditional receptor is called the \textit{reference molecule}, which is the same molecule provided to the input encoder for posterior sampling.

We investigated different levels of sampling variance through a setting called the \textit{variability factor}, denoted $\lambda_{var}$. This parameter scales the standard deviations used to sample the latent space:
\begin{equation}
    z' = \mu + \lambda_{var} \sigma z
\end{equation}

We also created a technique for controlling of the amount of bias towards the reference molecule. Instead of using either the posterior or prior distribution, we can sample distributions whose parameters are linearly interpolated between those of the prior and posterior according to a \textit{bias factor}, referred to as $\lambda_{bias}$ here:
\begin{align}
\begin{split}
    z' &= \mu_{interp} + \sigma_{interp} z \\
    \mu_{interp} &= \lambda_{bias} \mu_{post} + (1-\lambda_{bias}) \mu_{prior} \\
    \sigma_{interp} &= \lambda_{bias} \sigma_{post} + (1-\lambda_{bias}) \sigma_{prior}
\end{split}
\end{align}

For every sampling method that we evaluated, we generated 100 samples for each protein-ligand complex in the test set.

\subsection{Evaluation metrics}

We measured the validity, novelty, and uniqueness of the molecules generated from our model, which were defined as follows: A molecule is \textit{valid} if it consists of a single connected fragment and is able to be sanitized by RDkit, which checks valency constraints and attempts to Kekulize aromatic bonds. A molecule is \textit{novel} if its canonical SMILES string was not in the training set. A molecule is \textit{unique} if its canonical SMILES string was not generated already in the course of test evaluations. We also relaxed the internal bond lengths and angles of each generated molecule in the context of the binding site by Universal Force Field (UFF) minimization.\cite{rappe1992uff} We measured the internal energy and root-mean-squared-deviation (RMSD) of the molecules via UFF minimization. Both real and generated molecules then underwent Vina minimization and scoring with respect to the receptor. Lastly, we estimated the binding affinity of the minimized structures using an ensemble of CNN scoring functions that were trained on the CrossDocked2020 data set. The Vina minimization and CNN affinity prediction were performed using \texttt{gnina}, a deep learning-based molecular docking program.\cite{mcnutt2021gnina}

\begin{figure*}[t]
 \centering
 \includegraphics[width=\textwidth]{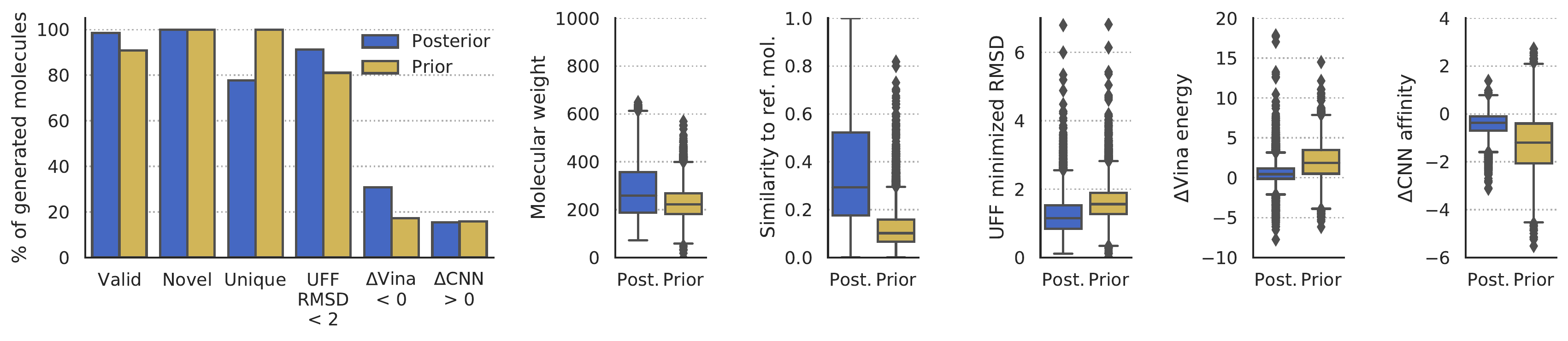}
 \caption{\textbf{Properties of generated molecules.} The percent of generated molecules that were valid, novel, unique, moved less than \SI{2}{\angstrom} RMSD during UFF minimization, had lower Vina energy, or had higher CNN predicted affinity than the reference molecule. These metrics are reported separately for molecules from posterior and prior sampling. Also shown are the distributions of molecular weight, Tanimoto fingerprint similarity, RMSD from UFF minimization, difference in Vina energy, and difference in CNN affinity. The fingerprint similarity, difference in Vina energy, and difference in CNN affinity were computed with respect to the reference molecule (lower $\Delta$Vina energy is better, higher $\Delta$CNN affinity is better).}
 \label{fig:post_prior_all_plots}
\end{figure*}

\section{Results}

\subsection{Properties of generated molecules}

\paragraph*{Validity, novelty, and uniqueness.} 98.5\% of molecules generated from posterior sampling were valid and 90.9\% generated from prior sampling were valid, as seen in Figure~\ref{fig:post_prior_all_plots}. 100\% of all generated molecules were novel, indicating that the model did not simply memorize the training set. Furthermore, 77.7\% of posterior molecules and 99.9\% of prior molecules were unique.

\paragraph*{Fingerprint similarity.} The distribution of Tanimoto fingerprint similarity in Figure~\ref{fig:post_prior_all_plots} shows that generated molecules tended to be quite dissimilar from the reference molecule. Posterior molecules had an average similarity of 0.33 to the reference molecule, though 25\% of posterior molecules had similarity greater than 0.5. Prior molecules were highly dissimilar from the reference molecule, with 25\% having similarity greater than 0.15.

\paragraph*{Per-target diversity.} Figure~\ref{fig:2D_div_box_plots} shows the diversity of generated molecules sampled from the same conditional receptor, measured as the inverse of their expected Tanimoto fingerprint similarity. Using a variability factor of 1.0, the per-target diversity was around 6 for posterior molecules and 5.5 for prior molecules. The diversity was significantly reduced when the variability factor was decreased from 1.0, and slightly reduced when it was increased.

\paragraph*{Shape similarity.} Two measures of molecular shape similarity are depicted in Figure~\ref{fig:shape_sim_box_plots}. The L2 loss from fitting atoms to real atomic density grids was close to zero, while it was in the 20-35 range for densities produced by the generative model. The shape similarity between generated molecules and reference molecules was also computed by RDKit. The shape similarity was around 0.62 for posterior molecules and 0.34 for prior molecules.

\paragraph*{Molecular weight and drug-likeness.} Prior molecules were smaller than posterior molecules by about \SI{50}{\dalton}, but there was considerable overlap in their molecular weight distributions, shown in Figure~\ref{fig:post_prior_all_plots}. A comparison of the quantitative estimate of drug-likeness (QED) score\cite{bickerton2012qed} between real and generated molecules is shown in Figure~\ref{fig:QED_box_plots}. Posterior molecules have a similar drug-likeness distribution as real molecules, while the QED score distribution for prior molecules has slightly heavier tails.

\paragraph*{UFF energy minimization.} When minimizing the energy of the generated molecules with UFF, we measured both the change in energy and the RMSD of the initial pose to the minimized pose. Figure~\ref{fig:QED_box_plots} compares the change in energy of generated molecules and real molecules. The energy decreased on the order of $-10^3$ kcal/mol and $-10^4$ kcal/mol for posterior and prior molecules, respectively, compared to $-10^2$ kcal/mol for real molecules. During UFF minimization, the conformation changed by less than \SI{2}{\angstrom} in 91.3\% of posterior molecules and 81.0\% of prior molecules.

\paragraph*{Vina energy and predicted binding affinity.} The relative stability of the generated molecules in the receptor binding site was quantified as the difference in Vina energy and CNN predicted binding affinity compared with the reference molecule, which was also minimized with Vina. The Vina energy and predicted affinity of posterior molecules were similar to those of the reference molecule, but shifted slightly towards higher energy and lower affinity. The Vina energy of prior molecules was significantly higher than reference ligands, and the predicted affinity tended to be lower and more variable. The diversity in the generated molecules was represented in their Vina energy and CNN affinity scores. Some structural differences decreased the stability and others improved it. 30.8\% of posterior molecules and 17.3\% of prior molecules had lower minimized Vina energy than the reference molecule. Moreover, 15.4\% of posterior molecules and 15.9\% of prior molecules had greater predicted affinity than the reference molecule after minimization. That is to say, a significant minority of sampled molecules were predicted to bind more strongly to the receptor than the reference ligand.

\paragraph*{Atom type distributions.} Figure~\ref{fig:gen_atom_properties} shows the distribution of atomic properties in the generated molecules, while Figure~\ref{fig:atom_properties} shows the atomic property distributions for real molecules from the CrossDocked2020 data set. The generative model produced diverse atom types that mostly matched the training set distribution, with a few exceptions. The model did not generate any boron, iron, bromine, or iodine atoms in our test evaluations, despite that these elements were in the training data. Additionally, the model rarely generated atoms with formal charges.

\paragraph*{Bond length distributions.} Figure~\ref{fig:bond_len_box_plots} compares the distributions of minimized bond lengths in real and generated molecules for the ten most common bond types. Overall, the bond lengths of real and generated molecules were fairly similar. The most noteworthy deviations were aromatic carbon-carbon and carbon-nitrogen bonds in generated molecules, which tend to be longer than in real molecules. The variance in length for these bonds was especially high in prior molecules.

\paragraph*{Bond angle distributions.} Figure~\ref{fig:bond_angle_box_plots} depicts the bond angle distributions for real and generated molecules after minimization. The median angle for many common bond angle types tended to be similar in real and generated molecules, but there was more variance in generated bond angles. Small, strained bond angles were fairly prevalent in generated molecules even after minimization. This is evidenced by the lower first quartile in a few of the bond angle distributions, in particular for carbon-oxygen-carbon.

\paragraph*{Torsion angle distributions.} Figure~\ref{fig:tors_angle_box_plots} portrays the distributions of torsion angles in minimized real and generated molecules. There are notable differences in the torsion angle distributions after UFF minimization. The distributions of torsion angles have different modes for generated molecules than real molecules for a number of common torsion angle types. On the other hand, aromatic torsion angles were centered at zero and had very low variance. This indicates that aromatic rings in generated molecules tended to be planar, as expected.

\begin{figure*}[t]
 \centering
 \includegraphics[width=\textwidth]{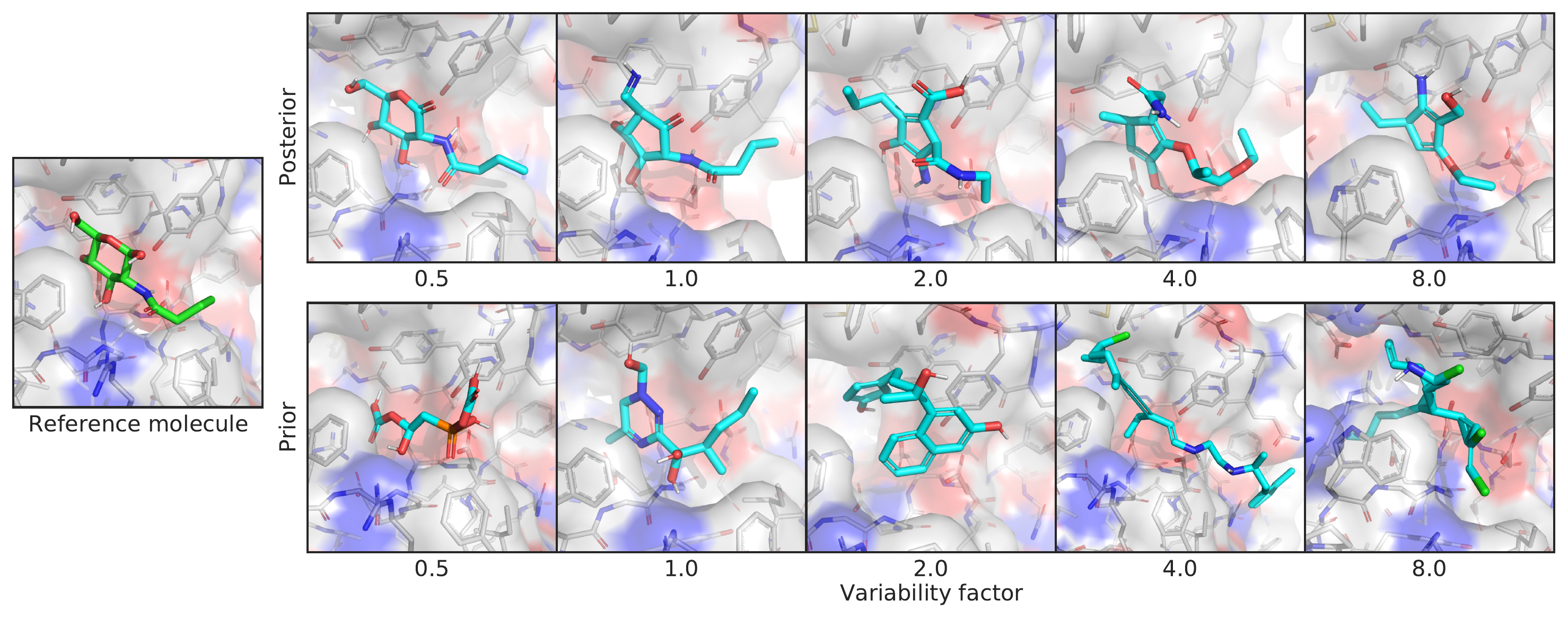}
 \caption{\textbf{Controlling the variability of generated molecules.} This figure depicts the effect of sampling molecules using different multipliers on the standard deviation of the latent distribution. The leftmost image shows the real ligand that was input to the model for posterior sampling. The first row shows posterior molecules sampled using different variability factors. The second row shows prior samples with different variability factors.}
 \label{fig:var_factor_structs}
\end{figure*}

\begin{figure*}
 \centering
 \includegraphics[width=\textwidth]{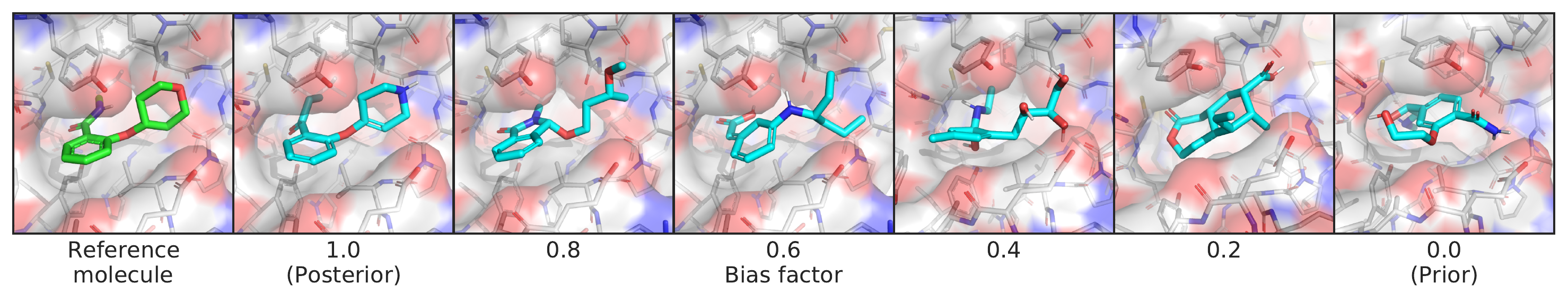}
 \caption{\textbf{Controlling bias towards the reference molecule.} This figure shows the effect of sampling molecules from latent distributions that interpolate between the posterior and prior. On the far left is the real molecule that was used to define the posterior distribution, followed by molecules sampled using different bias factors. A bias a factor of 1.0 indicates the full posterior distribution and 0.0 indicates the full prior distribution.}
 \label{fig:post_factor_structs}
\end{figure*}

\subsection{Controlling sampling variability and bias}

\paragraph*{Posterior variability.} We tested whether controlling the amount of sampling variability would alter the distribution of generated molecules in useful ways. Increasing the variability factor caused a corresponding increase in the diversity of the generated molecules, while decreasing this parameter reduced the diversity. Specifically, molecules generated from the posterior appeared more similar to the input ligand when using a lower variability factor. This is reflected in Figure~\ref{fig:var_factor_post_box_plots}, where lower variability factors produced posterior molecules with higher Tanimoto fingerprint similarity to the reference, and higher variability factors decreased the fingerprint similarity. An example of this relationship is shown in the first row of Figure~\ref{fig:var_factor_structs}, where the posterior molecule with the lowest variability factor is identical to the input ligand except for replacing an alcohol with a ketone. As the variability factor increased, more functional groups were modified and geometric changes were more drastic. At the highest variability factor, the molecule has only a faint scaffold similarity to the reference ligand. In addition to being more diverse, posterior molecules generated with a higher variability factor also tended to be less energetically stable and favorable in the binding pocket.

\paragraph*{Prior variability.} The effect of the variability factor on prior molecules, exemplified in the second row of Figure~\ref{fig:var_factor_structs}, was less straightforward. There was no relationship between the variability factor and similarity to the reference molecule, but there was a positive association with the average size and complexity of the molecules. This is evidenced by Figure~\ref{fig:var_factor_prior_box_plots}, which shows that the molecular weight and energy of prior molecules increased with the variability factor.  This can be explained by considering two facts: 1) increasing the variability decreases the expected probability of the generated samples under the learned prior distribution (and by extension, increases their distance from the training data manifold), and 2) neural networks with ReLU activation functions are piece-wise linear. The farther that we sample from the training data manifold, the more likely it is that we end up in a linear region of a network activation, any of which could be associated with the generated density magnitude in some location. This would explain why we observe larger molecules when sampling less probable regions of latent space. Although prior molecules with increased variability tended to be less favorable in terms of energy and binding affinity, the variance in these metrics increased as well. At higher variability factors, there were still a significant fraction of prior molecules that had lower Vina energy and higher predicted affinity than the reference molecule.

\begin{figure*}[t]
 \centering
 \includegraphics[width=\textwidth]{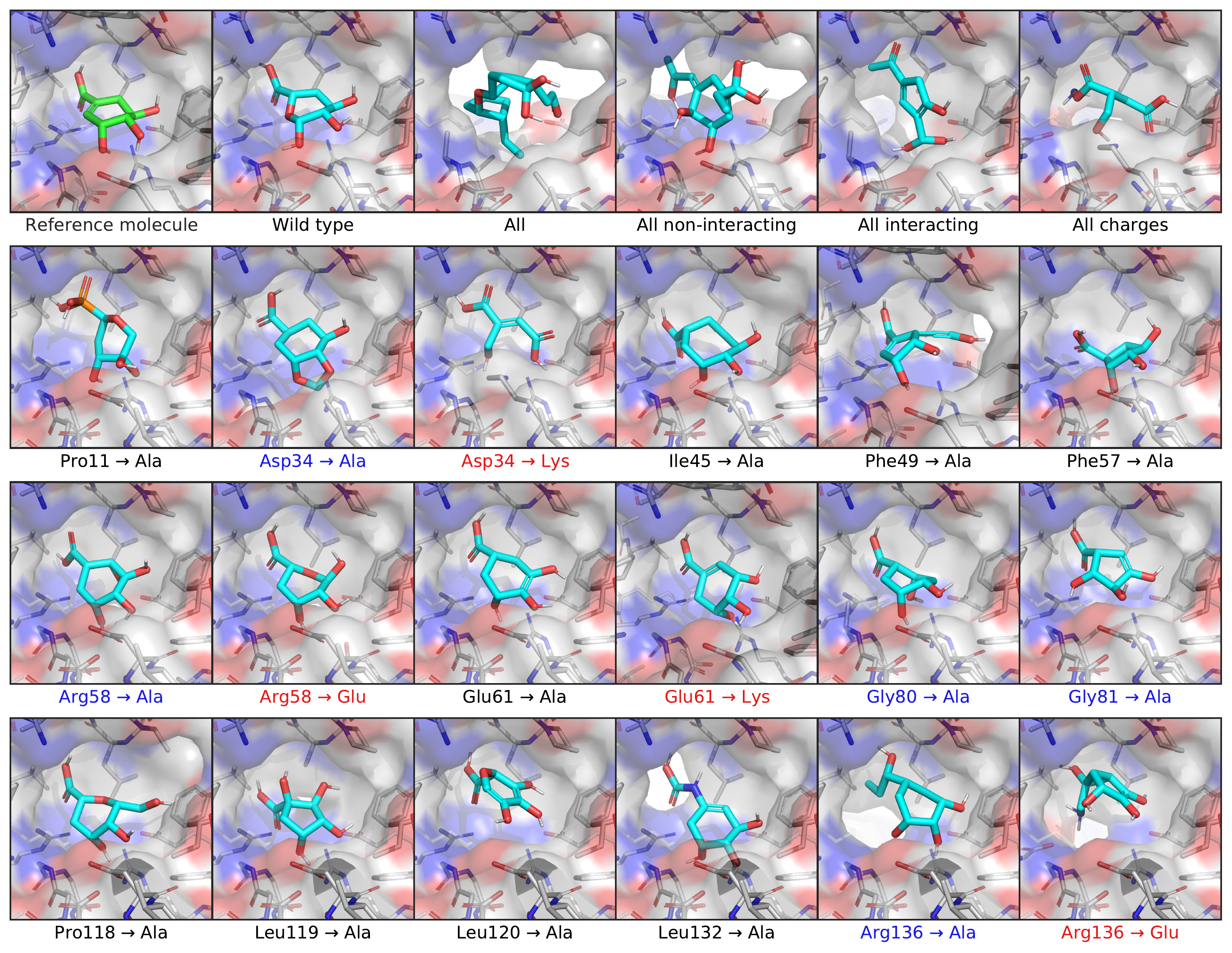}
 \caption{\textbf{Conditioning generated molecules on shikimate kinase mutants.} This figure displays posterior molecules that were generated using shikimate as the input ligand, shown in the top left corner. They were each conditioned on mutated versions of the shikimate kinase receptor. After the reference molecule, the first row shows molecules generated from the cognate receptor (Wild type) and four different multi-residue mutants. The next three rows show molecules conditioned on receptors with different single-residue mutations. The mutations highlighted in blue involve residues identified in previous work as making important binding interactions with shikimate. Mutations that inverted the charge of the residue are highlighted in red.}
 \label{fig:diff_cond_rec_structs}
\end{figure*}

\paragraph*{Bias towards reference molecule.} Next we evaluated the impact of sampling from interpolated latent distributions using the bias factor. Figure~\ref{fig:post_factor_structs} depicts a set of molecules generated using a single reference molecule with different bias factors, ranging from fully-posterior to fully-prior. The molecule with the highest bias factor is extremely similar to the reference ligand. As the sampled distribution became more prior-like, the molecules grew increasingly dissimilar from the reference ligand, converging on the distribution of prior molecules. This effect is shown quantitatively in Figure~\ref{fig:post_factor_box_plots}. As the bias factor increased, the molecular weight of the generated molecules approached the posterior distribution of the reference molecule, and the Tanimoto similarity grew as well. Increasing the bias factor also gradually shifted the distributions of minimized RMSD, difference in Vina energy, and difference in predicted affinity from that of the prior molecules to that of the posterior molecules.

\begin{figure*}[t]
 \centering
 \includegraphics[width=\textwidth]{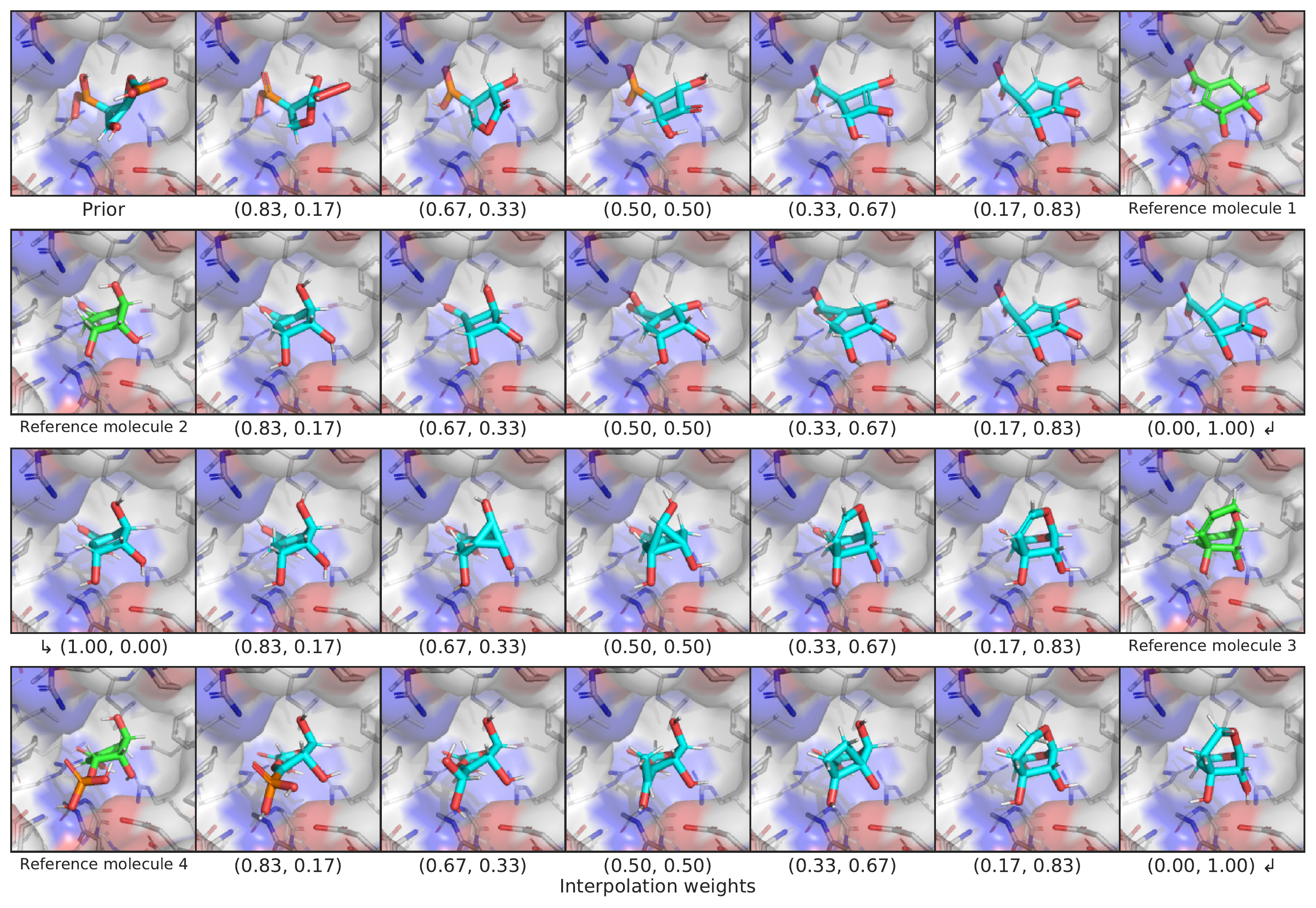}
 \caption{\textbf{Latent interpolation between shikimate kinase ligands.} This figure depicts a series of spherical interpolations in latent space between four different known actives for shikimate kinase. Starting with a prior molecule, each row displays an interpolation to the next ligand in the sequence, with the real molecule shown at the end of the row. The interpolated molecules are labelled with the weights that were used to combine the two endpoints of the latent interpolation. The molecules in this graphic were not minimized with any force field.}
 \label{fig:latent_interp_structs}
\end{figure*}

\subsection{Conditioning on mutated receptors}

To test the extent that our model uses the conditional receptor when generating molecules, we compared molecules generated from the same reference ligand, but conditioning the generative process on mutated versions of the receptor. Then we compared the molecular similarity, difference in Vina energy, and difference in CNN affinity with respect to the reference molecule. The Vina energy and predicted affinity were calculated using the conditional (i.e. mutant) receptor in this evaluation, for both real and generated molecules. Therefore, these metrics can be used to determine whether the model takes the conditional receptor structure into account or if it generates similar molecules regardless.

\paragraph*{Shikimate kinase target.} The target we selected for this analysis was shikimate kinase from \textit{mycobacterium tuberculosis} (PDB ID: 1zyu).\cite{gan2006shikimate} This enzyme is involved in the biosynthesis pathway for chorismate, a precursor for aromatic amino acids, in bacteria. Because it is not found in animals or plants, shikimate kinase has been proposed as a promising target for the development of non-toxic antimicrobial agents for a variety of uses including the treatment of tuberculosis.\cite{coracini2016sktarget} We focused on the shikimate binding site of the kinase in this work, which has been suggested to be more druggable than the ATP binding site. The residues that were identified in the literature as important for binding with shikimate are Asp34, Arg58, Glu61, Gly80, Gly81, and Arg136. Out of these, the charged residues Asp34, Arg58, and Arg136 were most frequently found to interact with the ligand. Asp34 and Arg58 were specifically depicted participating in hydrogen bonds.

\paragraph*{Creating mutant receptors.} We considered every residue of shikimate kinase within \SI{5}{\angstrom} of the docked pose of shikimate as a binding pocket residue. For each of these, we mutated the receptor by replacing the residue with alanine. For charged residues, we created additional mutants by replacing the residue with an oppositely charged residue of similar size. Finally, we created four additional structures with multi-residue mutations. One of these replaced every binding pocket residue with alanine (All). Another replaced every residue known to be relevant for binding shikimate with alanine (All interacting). A third replaced every binding pocket residue \textit{not} considered relevant for binding with alanine (All non-interacting). Finally, we created a mutant that replaced every charged binding pocket residue with an oppositely charged one of similar size (All charges).

\paragraph*{Conditioning on multi-residue mutants.} Molecules generated based on mutated shikimate kinase receptors are depicted in Figure~\ref{fig:diff_cond_rec_structs}. The molecule generated from the wild type receptor is quite similar to shikimate, except for breaking the double bond and inserting an oxygen in the ring. On the other end of the spectrum, replacing every binding pocket residue with alanine caused drastic changes to the generated molecule. The crucial carboxylic acid group was removed, the scaffold expanded, and the overall hydrophobicity increased. When only the non-interacting residues were mutated, the molecule expanded towards the top of the pocket where most of the mutations occurred, but the polar moeities towards the bottom of the pocket remained available for hydrogen bonds with Arg58 or Arg136. The top oxygen of the carboxylic acid also turned into a carbon, reflecting its newly hydrophobic environment. In contrast, mutating the interacting residues transformed the bottom oxygen of the carboxylic acid into a carbon instead, which makes sense given that it could no longer form hydrogen bonds with arginine. Additionally, the remaining oxygen reoriented towards the top of the pocket, which became relatively hydrophilic. Inverting every charged residue diminished the size of the molecular scaffold and added a nitrogen to the end of the carboxylic acid. This transformed it from a strong hydrogen bond acceptor to a potential donor that could interact with the glutamic acid that replaced Arg136.

\paragraph*{Conditioning on single-residue mutants.} The single-residue shikimate kinase mutations had varying effects on the generated molecules. Only a few mutations of the non-interacting residues caused notable effects. Replacing Pro11 with alanine transformed the carboxylic acid into a phosphate group. Changing Phe49 to alanine created a hole towards the right rear of the pocket where the model extended a rigid alcoholic tail. Transforming Leu132 to alanine extended the carboxylic acid into the new space by inserting a nitrogen atom. The rest of the non-interacting mutations caused only modest scaffold alterations, but mutating the interacting residues caused more interesting effects. Replacing Asp34 with alanine inserted a carbon to form a ring out of the two former hydroxyl groups, which was consistent with the reduced polarity and increased space at the front of the pocket. Replacing Asp34 with lysine broke the ring by removing a carbon, but kept the hydroxyl groups that might interact with the lysine. Converting Arg136 to alanine transformed the bottom oxygen of the carboxylic acid into an aliphatic tail extending into the open space. Mutating Arg136 to glutamic acid turned the carboxylic acid group from a hydrogen bond acceptor into a donor by adding a nitrogen. Surprisingly, none of the mutations to Arg58, Gly80, or Gly81 resulted in drastic chemical changes, despite that these were mentioned in past work as relevant for binding.

\paragraph*{Testing receptor conditionality.} To test whether conditioning on mutated receptors altered the generated molecules, we ran Kolmogorov-Smirnov tests on the distributions of fingerprint similarity, change in Vina energy, and change in CNN predicted binding affinity with respect to the reference molecule. The distributions we compared were for the posterior molecules conditioned on the mutant and the ones conditioned on the wild type receptor, which are shown in Figure~\ref{fig:mutate_cond_rec_1.0_box_plots}. We used a significance level of $\alpha=0.05$ with one-sided alternative hypotheses. We tested whether conditioning on the mutant \textit{decreased} the molecular similarity and change in Vina energy, and whether it \textit{increased} the change in predicted binding affinity for the conditional receptor.

For posterior molecules, the distributions of molecular similarity and change in predicted affinity were significantly different for each of the multi-residue mutations and several single-residue mutations. Modifying the interacting residue Arg136 caused significant differences in all three metrics. Mutating Asp34 to alanine significantly decreased the similarity and increased the change in predicted affinity, but did not decrease the change in Vina energy. Mutating Arg58 to alanine resulted in no significant differences. Interestingly, inverting the charge of either Arg58 or Asp34 only caused significant differences in the molecular similarity, but did not improve the change in Vina energy or predicted affinity. It was also unexpected that mutating the non-interacting residue Phe49 caused significant differences in all three metrics.

We tested for differences in the property distributions of prior molecules generated from mutant receptors, shown in Figure~\ref{fig:mutate_cond_rec_0.0_box_plots}, in the same way. There were no significant differences in molecular similarity, which was expected since prior molecules are not biased towards the reference molecule. However, there were significant improvements in change in Vina energy and CNN affinity for all multi-residue and Arg136 mutants. Replacing Asp34 or Arg58 with alanine caused significantly higher changes in predicted affinity, but not lower Vina energy. Inverting the charge of Asp34 or Arg58 lead to significantly lower change in Vina energy, but only increased the change in predicted affinity for Arg58.

\subsection{Latent space interpolation}

We explored the latent space of our model further through interpolation between different known ligands of shikimate kinase. Our test set contained seven different binders for this target, four of which were bound at the shikimate active site. We encoded each of these four ligands with our model to obtain posterior latent vectors, then performed a spherical interpolation in latent space passing through each latent vector, and decoded molecules along the resulting latent trajectory. Each latent vector was decoded using the conditional information from the shikimate kinase receptor, and the center of the conditional grids were interpolated smoothly between the real ligand centers. The resulting interpolation can be viewed in Figure~\ref{fig:latent_interp_structs}.

The initial prior molecule was quite dissimilar from the endpoint of the first interpolation (Reference molecule 1), but the intermediate steps resembled each endpoint. The molecule halfway between them has three hyxdroxyl groups bound to an aliphatic ring, but one phosphate was retained from the prior molecule in place of the carboxylic acid of Reference molecule 1. The next endpoint (Reference molecule 2) was extremely similar to the first, which was reflected in the interpolation between them. The conserved atoms smoothly moved through space except for an abrupt change in chirality at one of the ring carbons. A smooth translation of this carbon would have resulted in geometry with a different hybridization state and bond orders, so this sudden chiral shift better maintains the chemical similarity to the endpoints. The third endpoint (Reference molecule 3) had a slightly different scaffold, but similar functional groups. The trajectory passed through a few strained molecules with small rings that have high overall shape similarity. The final interpolation gradually transformed the newly added ring of Reference molecule 3 into an alkyl branch that then became a diol before finally ending at the phosphate group of Reference molecule 4.

\section{Discussion}

We have demonstrated for the first time the ability to generate three dimensional molecules conditional on receptor binding pockets with deep learning. Over 90\% of the generated molecules were valid, novel, and unique, though these metrics are insufficient to evaluate the quality of 3D molecular conformations. For this, we highlight the fact that over 80\% of the generated molecules moved less than \SI{2}{\angstrom} RMSD when minimized with UFF, which provides an indication of their energetic stability within the binding site. To further emphasize the potential utility of our model for discovering active molecules, a significant number of generated molecules had lower Vina energy and higher predicted binding affinity for the receptor than the reference molecule.

The generated molecules tended to have more strain than real molecules which is why we relaxed their internal bond lengths and angles with UFF. This is probably due to the lack of explicit bond information used by the model, which instead relies only on the relative positions of local density maxima to determine atom locations. Small differences in the generated density can alter the position of a single atom, which can translate into disproportionately high energy. One promising avenue for future work is to integrate an energetic term into the loss so that the model is directly trained to produce stable molecules. This would require the ability to differentiate through the atomic coordinates, which is easier with an atomistic representation than atomic density grids. We are currently exploring a multi-modal approach that combines density grids with 3D molecular graphs to learn both global shape and inter-atomic features.\cite{arcidiacono2021molucinate}

We successfully showed that our generative model conditions its output on the receptor structure. We qualitatively and statistically verified that the model produced chemically relevant modifications in the generated molecules when conditioned on shikimate kinase mutants. All multi-residue mutations caused significant changes in the properties of generated molecules, and some of the important single-residue mutations did as well. Modifying Arg136 caused significant changes in all relevant properties we assessed, for both posterior and prior molecules. There were some differences between what residues were reported in the literature as important for binding compared with the mutations that caused changes in our model's output. For instance, Arg58 mutations did not tend to cause significant changes, even though it was described as a hydrogen bond participant. Phe49 mutations had a significant impact on posterior molecules even though this residue was not described as interacting with shikimate. It was also surprising that inverting charges did not cause more drastic changes to the generated molecules. This could be due to inconsistent charges and protonation states in the training data.

We also found that the generation of molecules using atomic density grids is quite sensitive to the hyperparameters of the grid representation. Augmentation of training with random rotations was crucial for counteracting the coordinate-frame dependency of the density grids. There was also an interaction between the grid resolution and atomic radius used in the density kernel. When several atoms are nearby in the same grid channel (e.g. aromatic carbon rings), the density of each atom can overlap and produce a peak in the middle of the ring or bond. This makes it difficult to resolve the individual atoms through atom fitting. By reducing the radius of the density kernel, it becomes easier to distinguish atoms in close proximity at a given grid resolution, thereby producing more accurate and chemically realistic structures. Optimization of these grid settings had a significant impact on the quality of the generated molecules.

In future work, we will experiment with training setups that emphasize the use of the conditional receptor. One interesting augmentation would be to apply different random rotations on the input branch and conditional branch, then train the model to generate ligands in the conditional coordinate frame. This would encourage a coordinate frame-invariant latent space and enforce reliance on the structural characteristics of the conditional receptor to determine the generated ligand orientation. Another enhancement would be to train using different receptors within the same binding pocket cluster of the CrossDocked2020 set in the input and conditional branch. A more challenging future direction would be to provide higher-RMSD ligand poses as the input to the model and train using the lowest-RMSD pose as the label, essentially performing instantaneous minimization for docking.

We hope that this work accelerates the usage of 3D protein structure in molecular generative models. There is vast potential for further development of this approach. To enable the reproduction and extension of this work, we provide all code for this project at \url{https://github.com/mattragoza/liGAN}.

\section*{Author Contributions}
\textbf{Matthew Ragoza:} Conceptualization, Software, Resources, Data curation, Methodology, Investigation, Validation, Visualization, Formal Analysis, Writing - original draft, Project administration.\\
\textbf{Tomohide Masuda:} Conceptualization, Software, Investigation, Methodology, Resources, Validation, Writing – review \& editing.\\
\textbf{David Ryan Koes:} Conceptualization, Software, Resources, Investigation, Validation, Writing – review \& editing, Project administration, Funding acquisition.

\section*{Conflicts of interest}
There are no conflicts to declare.

\section*{Acknowledgements}
 This work is supported by R01GM108340 from the National Institute of General Medical Sciences, is supported in part by the University of Pittsburgh Center for Research Computing through the resources provided, and used the Extreme Science and Engineering Discovery Environment (XSEDE), which is supported by National Science Foundation grant number ACI-1548562 through the Bridges GPU-AI resource allocation TG-MCB190049.


\balance

\renewcommand\refname{References}
\bibliography{biblio} 
\bibliographystyle{rsc} 

\pagestyle{fancy}
\thispagestyle{plain}
\fancypagestyle{plain}{
\renewcommand{\headrulewidth}{0pt}
}

\makeFNbottom
\makeatletter
\renewcommand\LARGE{\@setfontsize\LARGE{15pt}{17}}
\renewcommand\Large{\@setfontsize\Large{12pt}{14}}
\renewcommand\large{\@setfontsize\large{10pt}{12}}
\renewcommand\footnotesize{\@setfontsize\footnotesize{7pt}{10}}
\makeatother

\renewcommand{\thefootnote}{\fnsymbol{footnote}}
\renewcommand\footnoterule{\vspace*{1pt}%
\color{cream}\hrule width 3.5in height 0.4pt \color{black}\vspace*{5pt}} 
\setcounter{secnumdepth}{5}

\makeatletter 
\renewcommand\@biblabel[1]{#1}            
\renewcommand\@makefntext[1]%
{\noindent\makebox[0pt][r]{\@thefnmark\,}#1}
\makeatother 

\renewcommand{\figurename}{\small{Fig.}~}
\sectionfont{\sffamily\Large}
\subsectionfont{\normalsize}
\subsubsectionfont{\bf}
\setstretch{1.125} 
\setlength{\skip\footins}{0.8cm}
\setlength{\footnotesep}{0.25cm}
\setlength{\jot}{10pt}
\titlespacing*{\section}{0pt}{4pt}{4pt}
\titlespacing*{\subsection}{0pt}{15pt}{1pt}

\twocolumn[
\begin{@twocolumnfalse}
{
    \vspace{35pt}
}
\par
    \vspace{1em}
    \sffamily
    
    \noindent\LARGE{\textbf{Supplementary Materials: Generating 3D Molecules Conditional on Receptor Binding Sites with Deep Generative Models}} \\

    \noindent\large{Matthew Ragoza,\textit{$^{a}$} Tomohide Masuda,\textit{$^{b}$} and David Ryan Koes\textit{$^{c}$}} \\

\end{@twocolumnfalse} \vspace{1.2cm}
]

\renewcommand*\rmdefault{bch}\normalfont\upshape
\rmfamily
\section*{}
\vspace{-1cm}

\nobalance
\renewcommand\thefigure{S\arabic{figure}}
\setcounter{figure}{0}

{
\small
\textit{$^{a}$~Intelligent Systems Program, University of Pittsburgh, Pittsburgh, PA, 15213. E-mail: mtr22@pitt.edu}\\
\textit{$^{b}$~Department of Computational and Systems Biology, University of Pittsburgh, Pittsburgh, PA, 15213. E-mail: tmasuda@pitt.edu}\\
\textit{$^{c}$~Department of Computational and Systems Biology, University of Pittsburgh, Pittsburgh, PA, 15213. E-mail: dkoes@pitt.edu}\\
}

\begin{figure}
 \centering
 \includegraphics[width=\columnwidth]{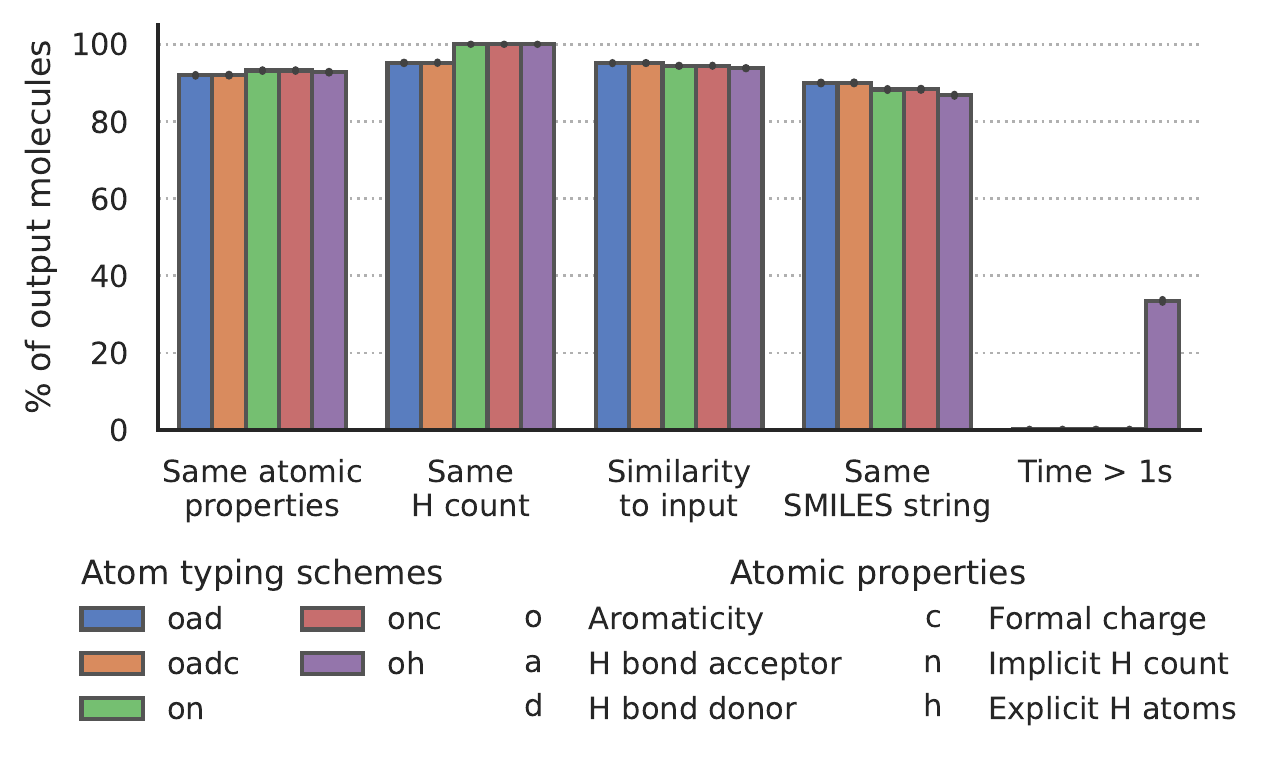}
 \caption{\textbf{Reconstructing molecules from atom types.} Ability to reconstruct real molecules from their atom types and coordinates through bond inference using different atom typing schemes. We selected ``oadc.''}
 \label{fig:bond_adding}
\end{figure}
\begin{figure*} 
 \centering
 \includegraphics[width=\textwidth]{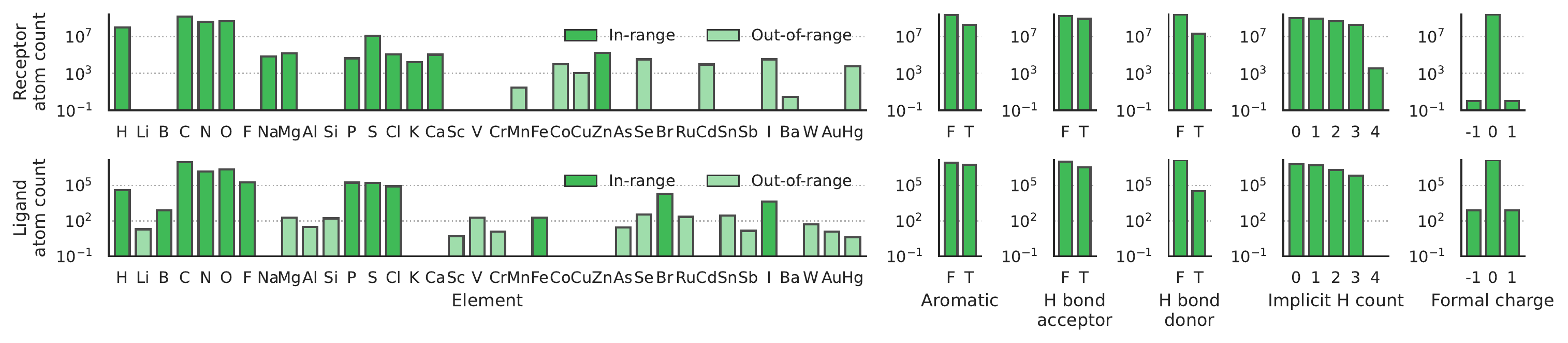} 
 \caption{\textbf{Atomic properties in the CrossDocked2020 data set.} Log-scale distributions of atomic properties in the CrossDocked2020 data set that were used to select value ranges represented in our atom type scheme (In-range). To limit the size of the density grids, rare elements were replaced with a placeholder element (Out-of-range). We used different elements for receptor and ligand atoms, but all other properties used the same ranges.}
 \label{fig:atom_properties}
\end{figure*}

\paragraph*{Atom typing scheme.} We enumerated the atoms in the CrossDocked2020 dataset and plotted their property distributions, seen in Figure~\ref{fig:atom_properties}, to create our atom typing scheme. Both for receptors and ligands, we considered element, formal charge, aromaticity, hydrogen bond acceptor/donor, and number of bonded hydrogens. We selected value ranges for the properties representative of the vast majority of the data set and used placeholders for out-of-range values. We also tried representing hydrogens explicitly. We evaluated atom typing schemes with different combinations of properties and selected the one that enabled the most accurate reconstruction of molecules. This evaluation is shown in Figure~\ref{fig:bond_adding}.

\paragraph*{Bond inference algorithm.} The goal of bond inference was to connect atoms into a single molecule with realistic bonds and  hydrogens subject to atom type constraints. We first added bonds between all atoms in within a distance range. Then we set the formal charge and hydrogen count based on the atom type. Any bonds between atoms with invalid valences were removed, as well as bonds that were excessively strained in length or angle. Next the hybridization states of aromatic atoms were set to sp2, the bonds between them were set as aromatic, and the orders of all bonds were perceived using OpenBabel. Finally, empty valences were filled with either hydrogens or higher bond orders.

\paragraph*{Atom fitting algorithm.} The atom fitting algorithm jointly optimized a set of atoms and their coordinates using a reference density through an iterative approach. In each iteration, the set of grid points with the highest density were evaluated as potential new atoms to expand the current structure. The atoms were individually added to the structure and their coordinates optimized by gradient descent with respect to the reference density. If adding the new atom decreased the loss (sum of squared error), the structure was stored. At the end of each iteration, the best new structure was set as the initial structure for the next iteration, and the remaining density was set as the new reference density. This repeated until no new atoms could be found that improved the loss.

\begin{algorithm}
\caption{Bond inference algorithm}\label{alg:bond_adding}
\KwData{$T \in \mathbb{R}^{N \times N_T}, C \in \mathbb{R}^{N \times 3}$}
\KwResult{$M \in $ Molecules}
$M \gets$ add all bonds within distance range($T, C$)\;
$M \gets$ add hydrogens and formal charges($M, T$)\;
$M \gets$ remove bad valences and geometry($M, T$)\;
$M \gets$ set hybridization and aromaticity state($M, T$)\;
$M \gets$ perceive bond orders based on geometry($M, T$)\;
$M \gets$ fill valences with hydrogen or bond orders($M, T$)\;
\end{algorithm}

\begin{algorithm}
\caption{Atom fitting algorithm}\label{alg:atom_fitting}
\KwData{$\mathbf{G}_{ref} \in \mathbb{R}^{N_T \times N_X \times N_Y \times N_Z}$}
\KwResult{$T \in \mathbb{R}^{N \times N_T}, C \in \mathbb{R}^{N \times 3}$}
$(T, C) \gets [], []$\;
\While{\textup{found new best structs}}{
    \ForEach{$(t_{new}, c_{new}) \gets$ \upshape rank points by density($\mathbf{G}_{ref}$)}{
        $(T_{new}, C_{new}) \gets ([T, t_{new}], [C, c_{new}])$\;
        $\mathbf{G}_{diff}, C_{new}, loss \gets $ gradient descent($\mathbf{G}_{ref}, T_{new}, C_{new}$)\;
        \If{$loss$ \textup{decreased}}{
            add $(loss, \mathbf{G}_{diff}, T_{new}, C_{new})$ to new best structs\;
        }
    }
    $(loss, \mathbf{G}_{ref}, T, C) \gets$ new best struct\;
}
\end{algorithm}

\begin{figure*} 
 \centering
 \includegraphics[width=0.45\textwidth]{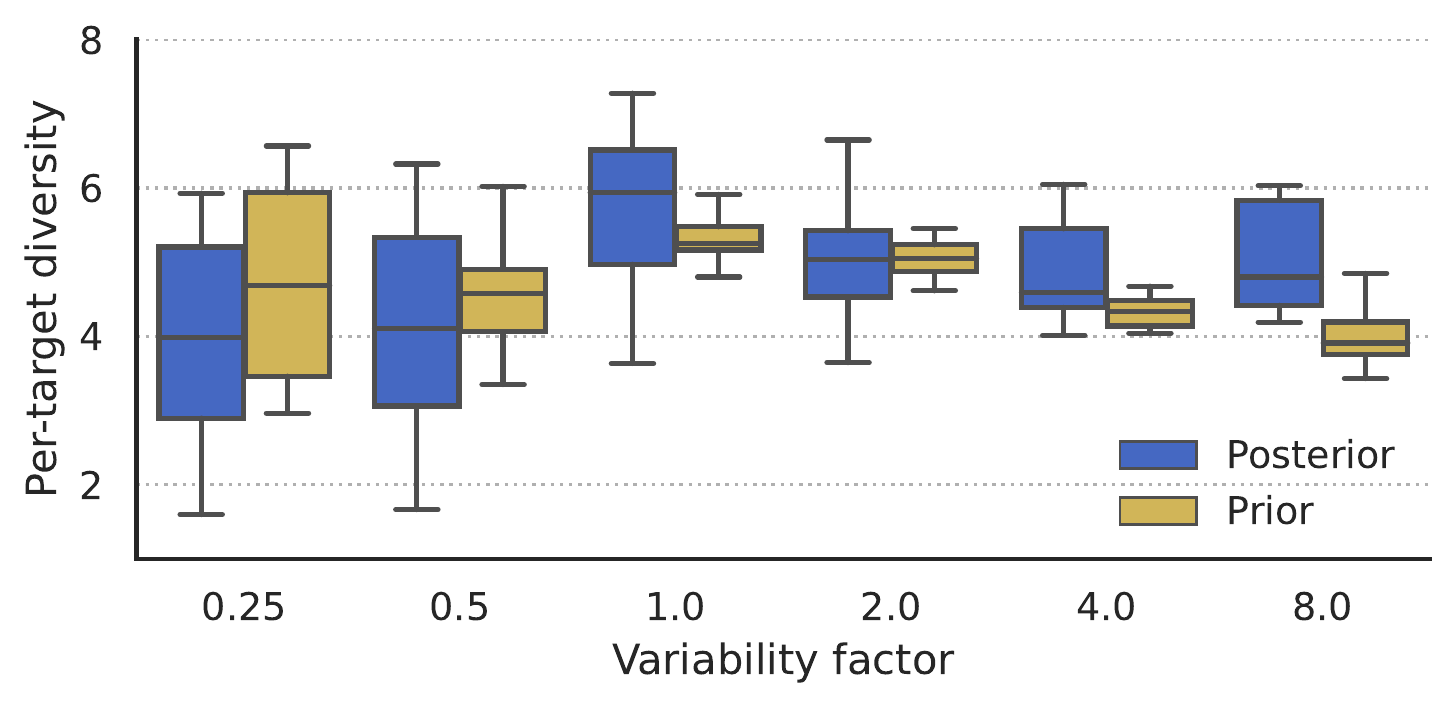}
 \caption{\textbf{Diversity of generated molecules.} Comparison of two-dimensional diversity of generated molecules for a given protein binding pocket with respect to the variability factor. The diversity was computed as the inverse of the expected Tanimoto fingerprint similarity of generated molecules conditioned on each receptor.}
 \label{fig:2D_div_box_plots}
\end{figure*}

\begin{figure*} 
 \centering
 \includegraphics[width=0.45\textwidth]{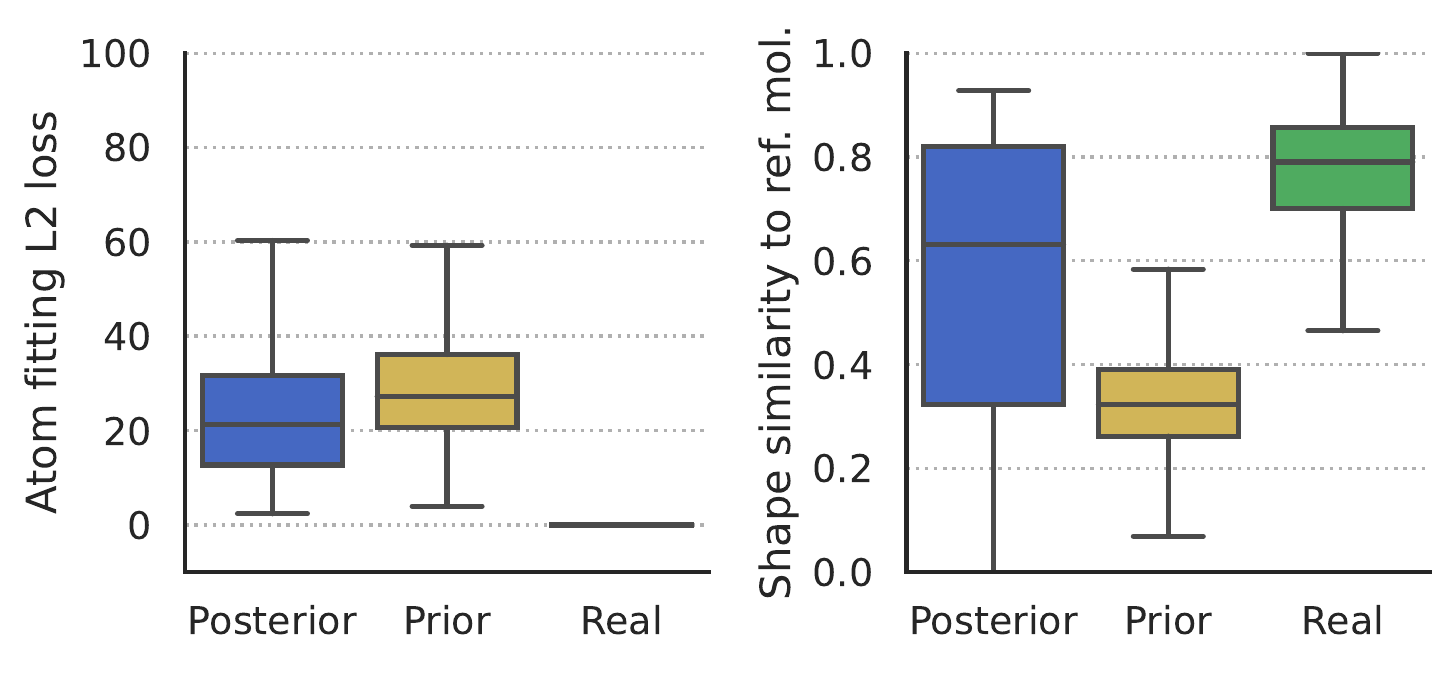}
 \caption{\textbf{Shape similarity metrics.} This figure shows shape similarity metrics for generated molecules. On the left is the L2 loss of the density representation of generated molecules with the generated density to which the molecules were fit using atom fitting (i.e. the objective function value minimized by atom fitting). On the right is the Tanimoto shape similarity between generated molecules and the reference molecule.}
 \label{fig:shape_sim_box_plots}
\end{figure*}

\begin{figure*} 
 \centering
 \includegraphics[width=0.45\textwidth]{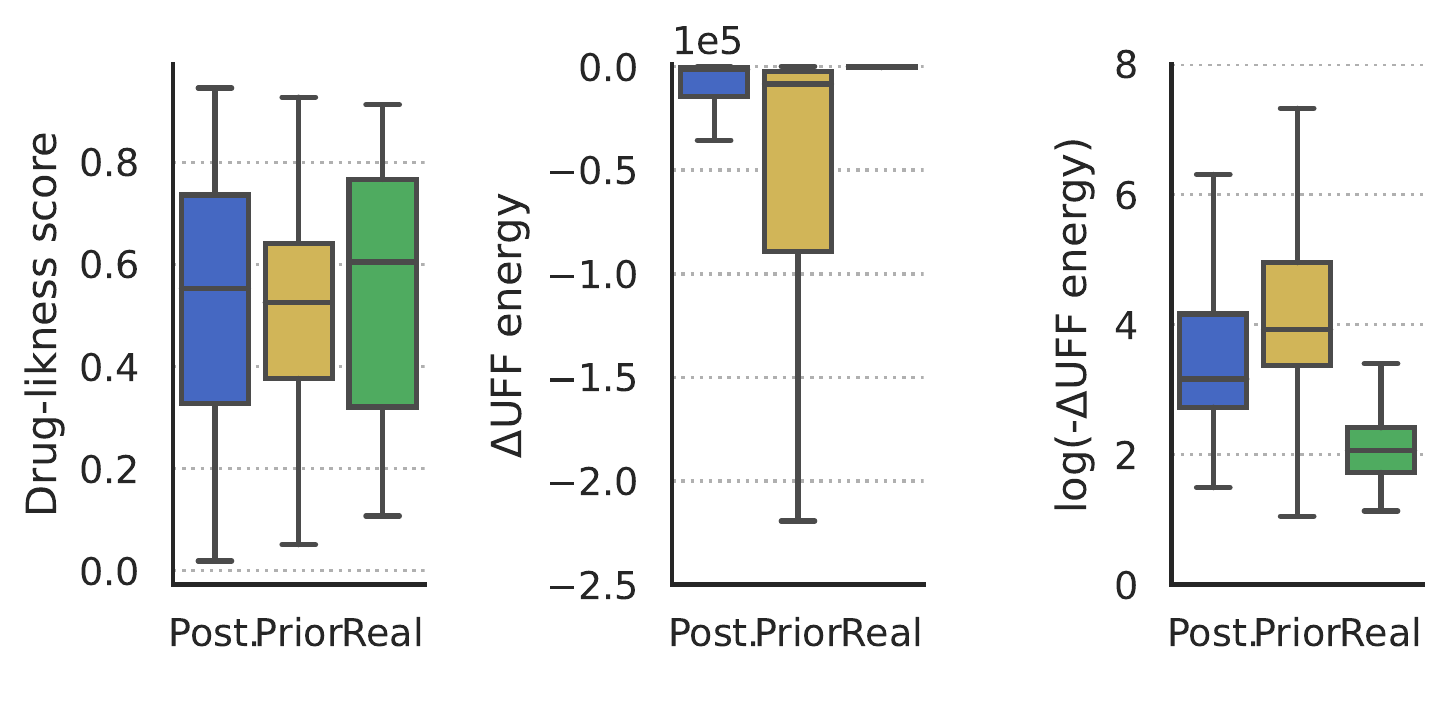}
 \caption{\textbf{Drug-likeness and change in internal energy.} This figure displays quantitative estimate of drug-likeness scores for generated and real molecules.\cite{bickerton2012qed} In addition, the change in energy due to UFF minimization is shown for real and generated molecules in linear and logarithmic scales.}
 \label{fig:QED_box_plots}
\end{figure*}

\newpage

\begin{figure*} 
 \centering
 \includegraphics[width=\textwidth]{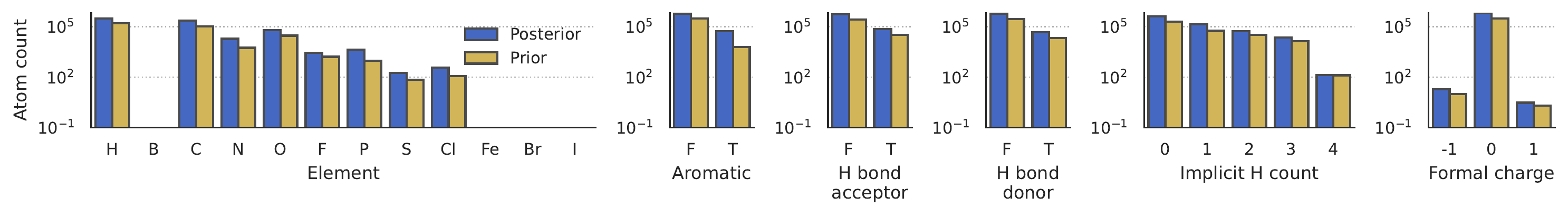} 
 \caption{\textbf{Atomic properties in generated molecules.} Log-scale distributions of atomic properties in molecules sampled from the generative model posterior and prior distributions.}
 \label{fig:gen_atom_properties}
\end{figure*}

\begin{figure*} 
 \centering
 \includegraphics[width=\textwidth]{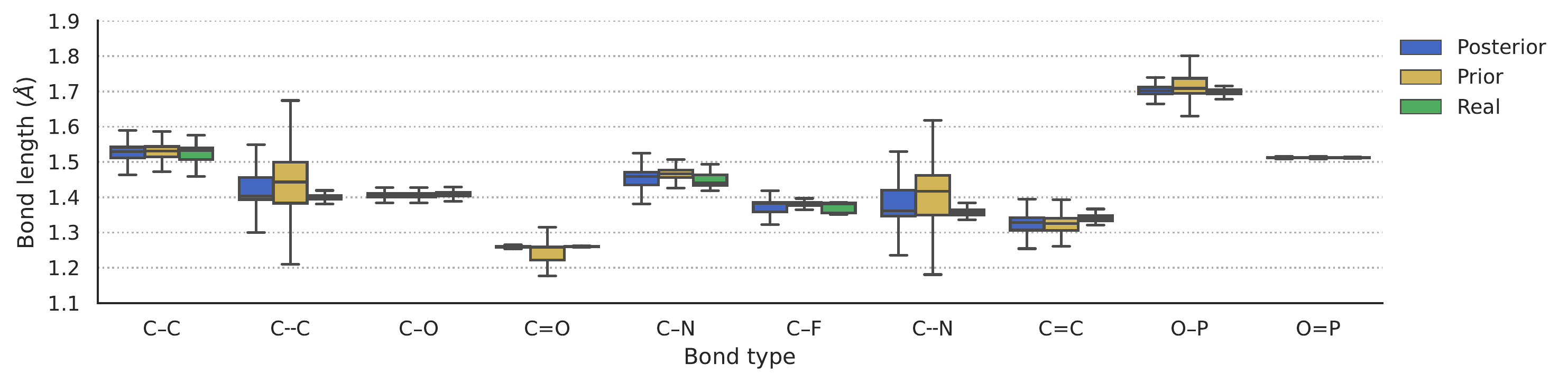}
 \caption{\textbf{Bond length distributions.} Comparison of distributions of bond lengths for the ten most common bond types in the data set. For both real and generated molecules, the bond lengths are shown after UFF minimization. The bond types are indexed by the elements, bond order, and aromaticity of the bonded atoms (aromatic bonds are indicated by a dashed line).}
 \label{fig:bond_len_box_plots}
\end{figure*}

\begin{figure*} 
 \centering
 \includegraphics[width=\textwidth]{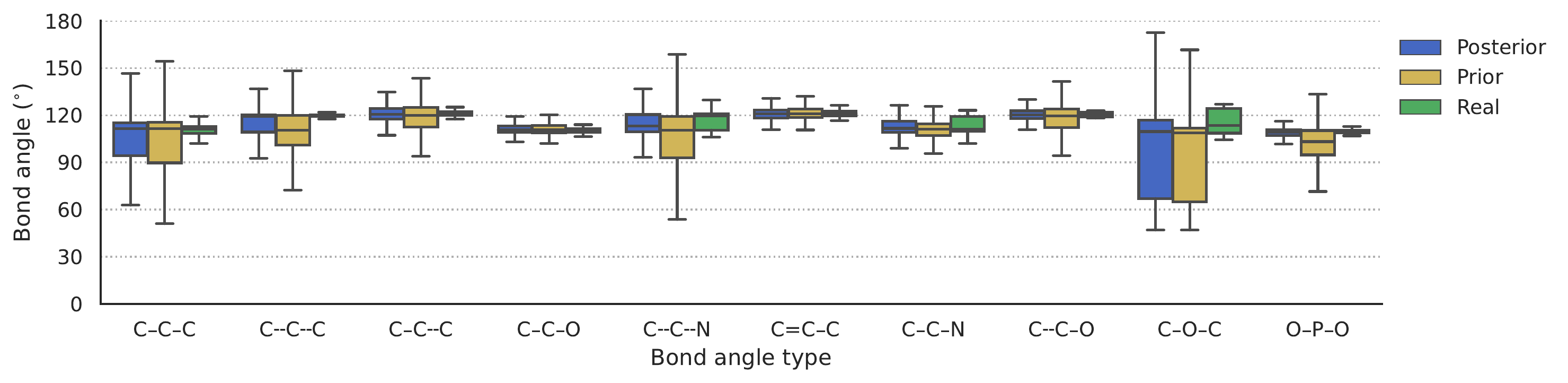}
 \caption{\textbf{Bond angle distributions.} Comparison of distributions of bond angles for the ten most common bond angle types in the data set. For both real and generated molecules, the bond angles are shown after UFF minimization. The bond angle types are indexed by the elements, bond order, and aromaticity of the bonded atoms (aromatic bonds are indicated by a dashed line).}
 \label{fig:bond_angle_box_plots}
\end{figure*}

\begin{figure*} 
 \centering
 \includegraphics[width=\textwidth]{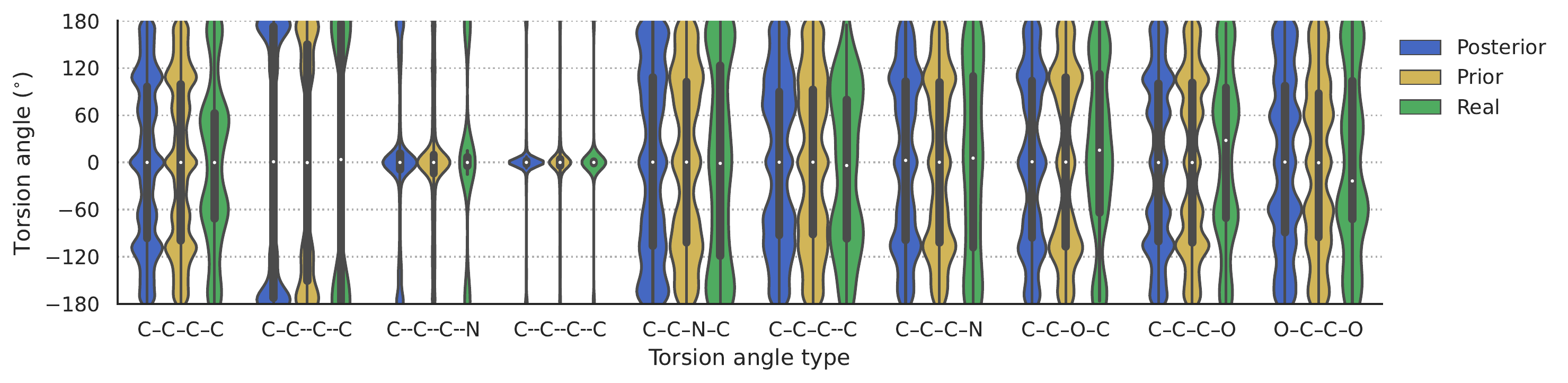}
 \caption{\textbf{Torsion angle distributions.} Comparison of distributions of torsion angles for the ten most common torsion angle types in the data set. For both real and generated molecules, the torsion angles are shown after UFF minimization. The torsion angle types are indexed by the elements, bond order, and aromaticity of the bonded atoms (aromatic bonds are indicated by a dashed line).}
 \label{fig:tors_angle_box_plots}
\end{figure*}

\begin{figure*} 
 \centering
 \includegraphics[width=\textwidth]{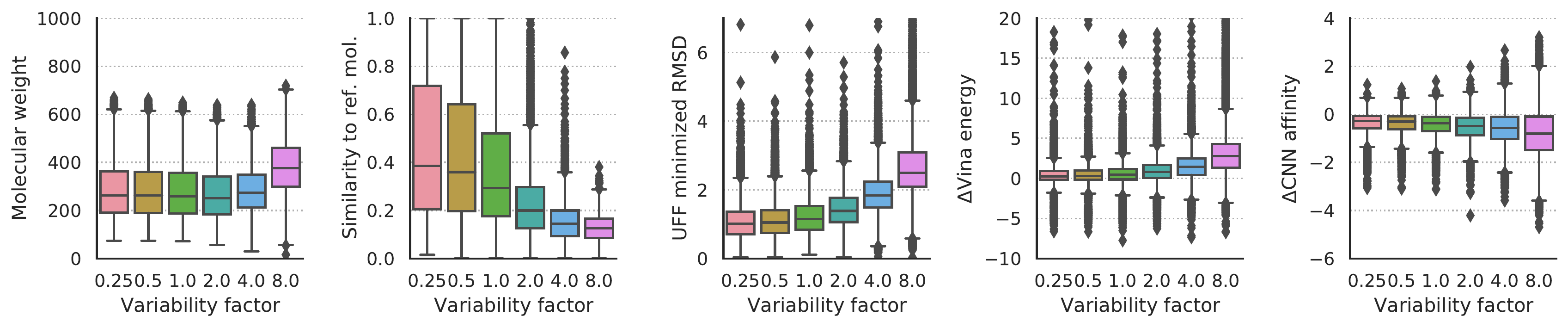}
 \caption{\textbf{Controlling variability of posterior molecules.} Properties of molecules generated from the posterior with different variability factors.}
 \label{fig:var_factor_post_box_plots}
\end{figure*}

\begin{figure*} 
 \centering
 \includegraphics[width=\textwidth]{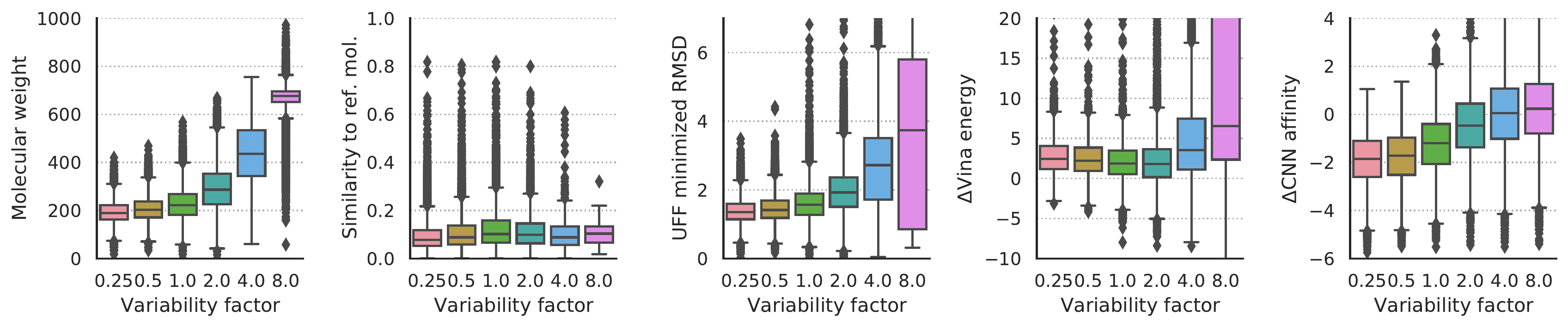}
 \caption{\textbf{Controlling variability of prior molecules.} Properties of molecules generated from the prior distribution with different variability factors.}
 \label{fig:var_factor_prior_box_plots}
\end{figure*}

\begin{figure*} 
 \centering
 \includegraphics[width=\textwidth]{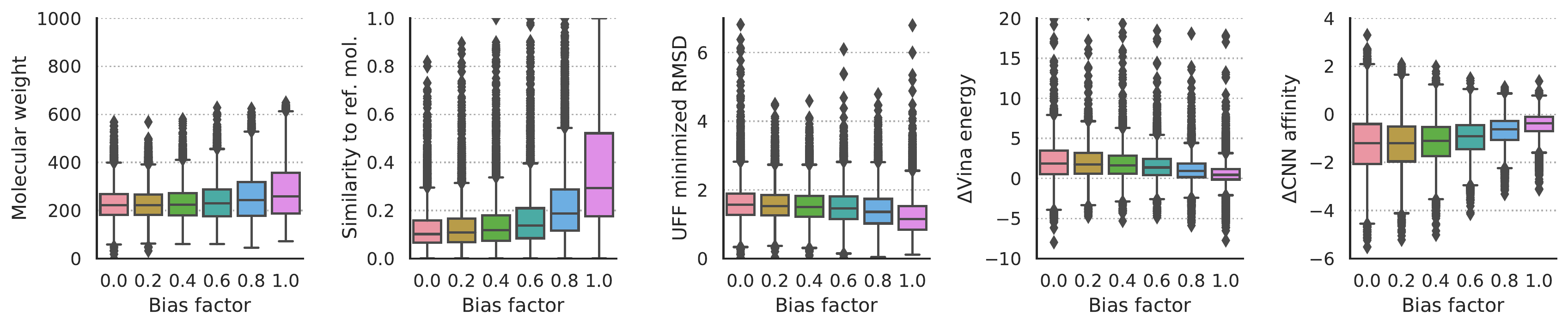}
 \caption{\textbf{Controlling bias towards reference molecule.} Properties of molecules generated from distributions that were interpolated between the prior and posterior. The prior distribution corresponds to a bias factor of 0.0 and the posterior distribution corresponds to a bias factor of 1.0.}
 \label{fig:post_factor_box_plots}
\end{figure*}

\begin{figure*} 
 \centering
 \includegraphics[width=\textwidth]{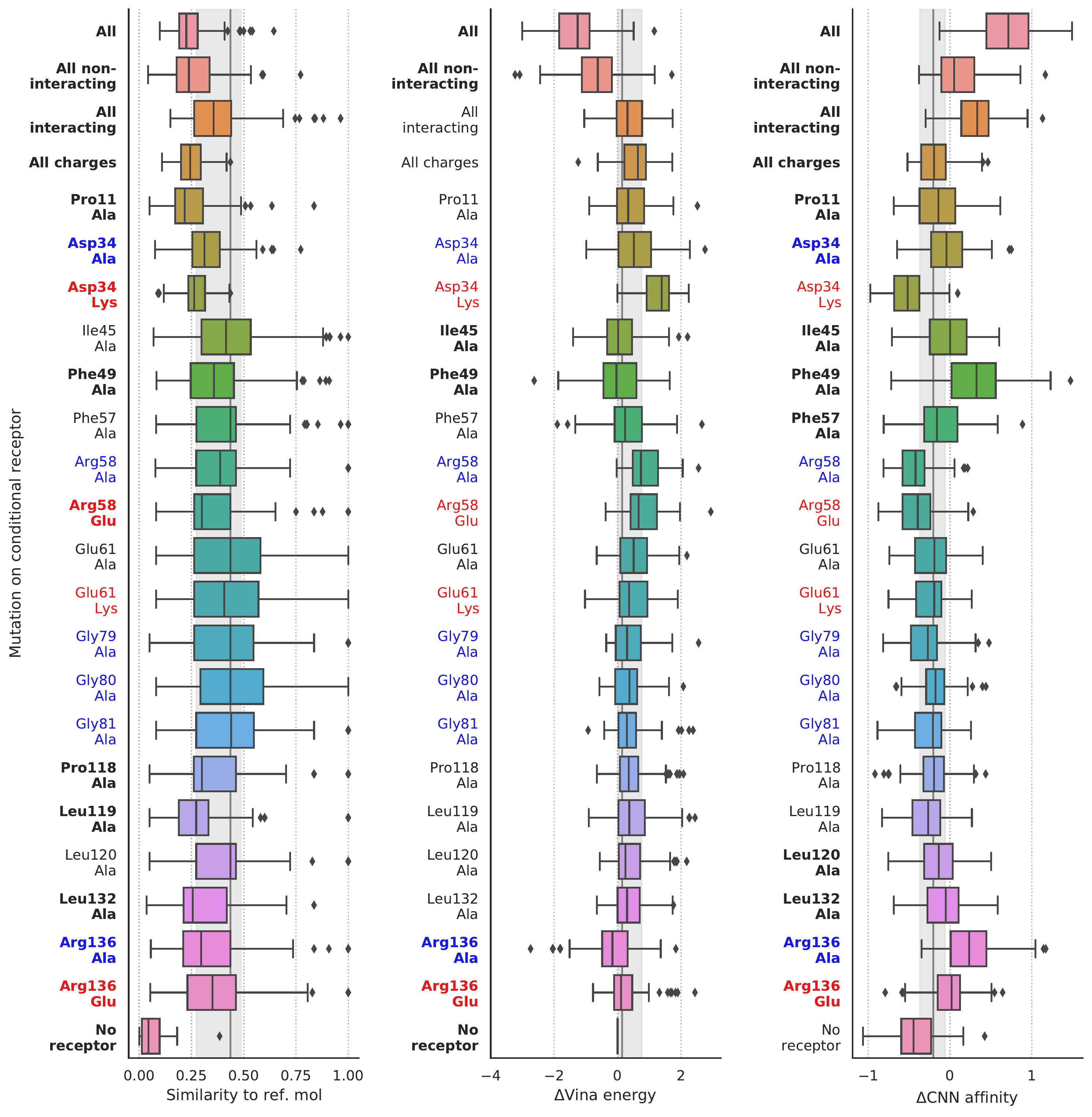}
 \caption{\textbf{Conditioning posterior molecules on mutant receptors.} Properties of molecules generated from the posterior distribution when conditioned on mutant receptors. Vina energy and CNN affinity were computed with respect to the conditional (mutant) receptor. Mutated residues highlighted in blue were described in past work as interacting with the known ligand. Mutations highlighted in red inverted the charge of the residue. Gray lines in the background show the property distribution (1st, 2nd, 3rd quartile) for molecules conditioned on the wild type receptor. Mutations that caused significantly different property distributions compared to the wild type receptor are shown in bold (one-sided Kolmogorov-Smirnov test with $\alpha = 0.05$).}
 \label{fig:mutate_cond_rec_1.0_box_plots}
\end{figure*}

\begin{figure*} 
 \centering
 \includegraphics[width=\textwidth]{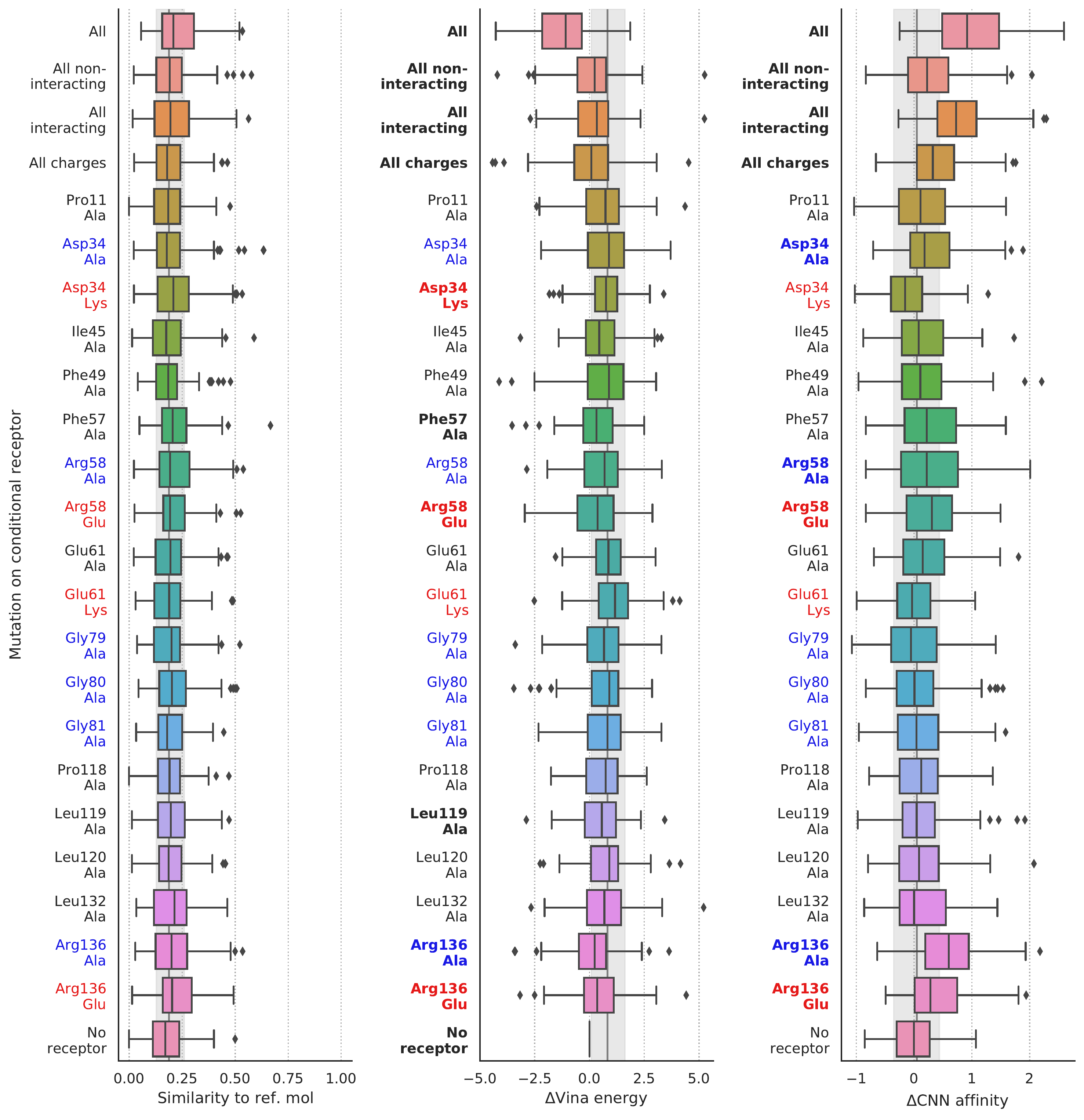}
 \caption{\textbf{Conditioning prior molecules on mutant receptors.} Properties of molecules generated from the prior distribution when conditioned on mutant receptors. Vina energy and CNN affinity were computed with respect to the conditional (mutant) receptor. Mutated residues highlighted in blue were described in past work as interacting with the known ligand. Mutations highlighted in red inverted the charge of the residue. Gray lines in the background show the property distribution (1st, 2nd, 3rd quartile) for molecules conditioned on the wild type receptor. Mutations that caused significantly different property distributions compared to the wild type receptor are shown in bold (one-sided Kolmogorov-Smirnov test with $\alpha = 0.05$).}
 \label{fig:mutate_cond_rec_0.0_box_plots}
\end{figure*}

\end{document}